# Moving Media as Photonic Heat Engine and Pump

Yoichiro Tsurimaki, Renwen Yu, and Shanhui Fan

Department of Electrical Engineering, Ginzton Laboratory, Stanford University, Stanford, 94305, CA

**Abstract**

A system consisting of two slabs with different temperatures can exhibit a non-equilibrium lateral Casimir force on either one of the slabs when Lorentz reciprocity is broken in at least one of the slabs. This system constitutes a photonic heat engine that converts radiative heat into work done by the non-equilibrium lateral Casimir force. Reversely, by sliding two slabs at a sufficiently high relative velocity, heat is pumped from the slab at a lower temperature to the other one at a higher temperature. Hence the system operates as a photonic heat pump. In this work, we study the thermodynamic performance of such a photonic heat engine and pump via the fluctuational electrodynamics formalism. The propulsion force due to the non-reciprocity and the drag force due to the Doppler effect was revealed as the physical mechanism behind the heat engine. We also show that in the case of the heat pump, the use of nonreciprocal materials can help reduce the required velocity. We present an ideal material dispersion to reach the Carnot efficiency limit. Furthermore, we derive a relativistic version of the thermodynamic efficiency for our heat engine and prove that it is bounded by the Carnot efficiency that is independent of the frame of reference. Our work serves as a conceptual guide for the realization of photonic heat engines based on fluctuating electromagnetic fields and relativistic thermodynamics and shows the important role of electromagnetic non-reciprocity in operating them.

## 1. **INTRODUCTION**

The effects of radiative heat transfer and non-equilibrium Casimir force can both occur between two objects having different temperatures, due to the exchange of energy and momentum as carried by thermally fluctuating electromagnetic fields emitted from the objects. The significant enhancement of these effects [1-20], when the two objects are brought in close proximity separated by nanoscale gaps, have motivated extensive studies due to the fundamental importance of these

E-mail correspondence: shanhui@stanford.edu

effects as well as opportunities for energy [21-25] and optomechanical applications at nanoscales [26-29]. The focus of most of these studies was on either radiative heat transfer or Casimir force alone. On the other hand, since both effects arise from the same underlying thermal electromagnetic fields, it should be of interest to study the relationship and the conversion between radiative heat transfer and non-equilibrium Casimir forces.

Recently, it was shown that materials with different temperatures can experience non-equilibrium Casimir force in the direction parallel to the interacting surfaces when at least one of them breaks Lorentz reciprocity, i.e., when at least one of them is made of non-reciprocal materials [30-37]. While lateral Casimir force can also exist in reciprocal systems both at thermal equilibrium [27,28,38-41] and non-equilibrium [42,43], the objects must break translational or rotational symmetry. Moreover, the lateral force in reciprocal systems both in thermal equilibrium and non-equilibrium results in a relaxation process to mechanical equilibrium, and continuous external stimuli such as external illumination of light or mechanical work are necessary to realize the persistent lateral force. In contrast, breaking Lorentz reciprocity of the system allows persistent non-equilibrium lateral Casimir force on translationally or rotationally invariant objects so far as the objects are in thermal non-equilibrium.

This existence of such a persistent lateral force suggests the possibility of a heat engine that converts the radiative heat transfer into mechanical work done by non-equilibrium lateral Casimir force. Recently, a single gyrotropic sphere at a different temperature from the environment was analyzed as a heat engine where the radiative heat transfer between the sphere and the environment results in the mechanical torque on the sphere [36]. The thermodynamic analysis revealed that the thermodynamic efficiency is bounded by the Carnot efficiency limit and magnetic hyperbolic plasma was proposed as a material that can be used to approach the Carnot limit in this system. The non-equilibrium lateral Casimir force in a parallel plate configuration has also been explored for the construction of a heat engine. [37]. The parallel plate geometry is desirable since it allows for the possibility of scaling up such a heat engine. The analysis in Ref. [37] however was based on the linear response theory of radiative heat transfer and non-equilibrium lateral Casimir force [44]. The validity of such linear response theory is limited to small temperature differences as well as small velocities of relative motion. To our best knowledge, the detailed analysis of two semi-infinite parallel slabs at relative motion as a heat engine has not been performed. Moreover, the reverse process, i.e., operation as a heat pump, was not analyzed before due to the requirement of high velocities of relative motion, which is beyond the applicability of the linear response scheme.

In this work, we analyze a system consisting of two semi-infinite parallel slabs in thermal non-equilibrium as a heat engine as well as a heat pump. We derive the fluctuational electrodynamics formalism of radiative heat transfer and non-equilibrium lateral Casimir force for anisotropic slabs at a relative motion with arbitrary velocities. At low velocities of relative motion, we show that the system of two slabs in thermal non-equilibrium works as a heat engine when at least one slab breaks Lorentz reciprocity. We reveal the physical mechanism of the conversion of radiative heat into mechanical work and how it evolves for different velocities of relative motion from the start of the heat engine to the steady state. We also analyze an ideal dispersion of materials that achieves Carnot efficiency. At high velocities of relative motion beyond the steady-state velocity, we show the operation of the system as a heat pump where the radiative heat flows from the slab at a lower proper temperature to the slab at a high proper temperature as a result of external work. In principle, we can conceive of operating the heat engine and pump at relativistic velocities. For such situations, we derive the relativistic thermodynamic efficiency and show that it is bounded by the Carnot efficiency that is independent of the frame of reference.

The paper is organized as follows. In Section II, we provide the fluctuational electrodynamics formalism that computes radiative heat transfer and lateral Casimir force between two semi-infinite slabs separated by a vacuum gap. We also discuss the constraints related to the second law of thermodynamics and imposed by Lorentz reciprocity. In Sections III and IV, we analyze the system as a heat engine and a heat pump, respectively. In Section V, we derive the relativistic thermodynamic efficiency of our heat engine. In Section VI, we summarize our findings.

## 2. FORMALISM

We consider two semi-infinite parallel slabs labeled as 1 and 2 that are separated by a vacuum gap $d$ as shown in Fig. 1. Two slabs can move laterally at a constant velocity $V$. We derive the radiative heat flux and lateral Casimir force per unit surface area, i.e., shear stress, between the two slabs in relative motion in the framework of fluctuational electrodynamics [45]. Previous studies have investigated radiative heat transfer between two slabs of anisotropic media at rest [46,47], shear stress on two slabs of isotropic media in a relative motion [48,49], and the friction coefficient, i.e., the ratio of the shear stress to the velocity of relative motion, in the linear response regime between two slabs of anisotropic materials [37]. As a step further, in this work we develop the fluctuational electrodynamic formalism for radiative heat flux and shear stress between two slabs of anisotropic materials in relative motion at arbitrary velocity. This formalism is applicable to anisotropic materials including those breaking Lorentz reciprocity. Our development is motivated by the observation that in the two-slab system, one of the slabs must be non-reciprocal in order to construct a heat engine driven by the non-equilibrium lateral Casimir force [36,37]. Moreover, to have a complete picture of the operation of such a heat engine, it is important to go beyond the linear response regime. In this section, we describe the main results of our formalism. A detailed derivation can be found in Supplementary Information (SI).

Due to quantum and thermal fluctuations, any object made of lossy materials emits electromagnetic waves. Radiative heat transfer and Casimir force between the slabs occur by such emission from one slab and the absorption by the other slab. The absorption and emission processes, moreover, involve multiple reflections of the waves between the slabs. The strength of the fluctuations, as well as the absorption and emission processes, depend on the temperatures and optical properties of the slabs. Since the physical properties of a slab are well defined in its rest frame where the local thermodynamic equilibrium is established, we incorporate the effects of the relative motion between the slabs by the Lorentz transformation of the electromagnetic waves in the vacuum between the slabs. Thus, we need not perform the Lorentz transformation of thermodynamic quantities.

In this work, we assume that slab 1 is at rest whereas slab 2 can move, and we refer to the frames in which slabs 1 and 2 are at rest as the rest frame and the co-moving frame, respectively. We use primes on the physical quantities in the co-moving frame. We consider the two slabs at the proper temperatures $T_1$ and $T_2'$, respectively. We assume the slabs are made of linear and nonmagnetic materials, i.e., the relative magnetic permeability is unity. Under the framework of fluctuational electrodynamics with these assumptions, our formalism is exact. Without loss of generality, we also assume slab 2 is moving at the velocity $V$ in the $x$-direction with respect to slab 1. For the observer in the rest frame, the net radiative heat flux $\varphi_{1\rightarrow 2}$ from slabs 1 to 2 and the net shear stress $f_{x,1}$ on slab 1 are given as

$$\begin{bmatrix} \varphi_{1\to2} \\ f_{x,1} \end{bmatrix} = \int_0^\infty \frac{d\omega}{2\pi} \int \frac{d\boldsymbol{q}}{(2\pi)^2} \begin{bmatrix} \hbar\omega \\ -\hbar q_x \end{bmatrix} [n_B(\omega, T_1) - n_B(\omega', T_2')] \tau_{1\to2}(\omega, \boldsymbol{q}; V), \quad (1)$$

where $\omega$ and $\boldsymbol{q} = (q_x, q_y)$ are the angular frequency and the in-plane wavevector components of the electromagnetic waves, $n_B(\omega, T) = \frac{1}{\exp\left(\frac{\hbar\omega}{k_B T}\right) - 1}$ is the Bose-Einstein distribution, and $\hbar$ and $k_B$ are the reduced Planck constant and the Boltzmann constant, respectively. $\tau_{1\to2}$ is the transmission coefficient for the electromagnetic waves emitted from slab 1 and absorbed by slab 2. For the propagative waves ($q = |\boldsymbol{q}| < k_0 = \frac{\omega}{c}$) and evanescent waves ($q > k_0$), the transmission coefficients are given as

$$\tau_{1\to2}(\omega, \boldsymbol{q}; V) = \begin{cases} \mathrm{Tr}\left[\left(I - \tilde{R}_2^\dagger \tilde{R}_2\right) \frac{1}{D_{12}} \left(I - R_1 R_1^\dagger\right) \frac{1}{D_{12}^\dagger}\right], & q < k_0 = \frac{\omega}{c} \\ \mathrm{Tr}\left[\left(\tilde{R}_2^\dagger - \tilde{R}_2\right) \frac{1}{D_{12}} \left(R_1 - R_1^\dagger\right) \frac{1}{D_{12}^\dagger} e^{-2\kappa_z d}\right] & q > k_0 \end{cases}, \quad (2)$$

where $I$ is the 2 by 2 identity matrix, $D_{12} = I - R_1 \tilde{R}_2 e^{2ik_z d}$, $k_z$ is the $z$-component of the wavevector in the vacuum, $\kappa_z = \sqrt{q^2 - k_0^2}$, and $\tilde{R}_2 = L R_2' (L^T)^{-1}$. $R_1$ and $R_2'$ are the reflection matrices of slabs 1 and 2 in the linear polarization basis, respectively, and their explicit expressions are

$$R_1 = \begin{bmatrix} r_1^{ss}(\omega, \boldsymbol{q}) & r_1^{sp}(\omega, \boldsymbol{q}) \\ r_1^{ps}(\omega, \boldsymbol{q}) & r_1^{pp}(\omega, \boldsymbol{q}) \end{bmatrix},$$
$$R_2' = \begin{bmatrix} r_2^{ss}(\omega', \boldsymbol{q}') & r_2^{sp}(\omega', \boldsymbol{q}') \\ r_2^{ps}(\omega', \boldsymbol{q}') & r_2^{pp}(\omega', \boldsymbol{q}') \end{bmatrix}, \quad (3)$$

where $r_k^{ij}$ ($i, j = s, p$ and $k = 1,2$) is the Fresnel reflection coefficient for the $j$-polarized incident light and the $i$-polarized reflected light for slab $k$. Note that the reflection matrix of slab 2, $R_2'$, is evaluated in the co-moving frame. The angular frequency and wavevector in the two frames are related by the Lorentz transformation as

$$\omega' = \gamma(\omega - q_x V),\ \ q_x' = \gamma(q_x - \beta k_0),\ \ q_y' = q_y,\ \ k_z' = k_z, \tag{4}$$

where $\beta = \frac{V}{c}$ and $\gamma^{-1} = \sqrt{1-\beta^2}$.

The matrix $L$ transforms the electric fields of forward propagating waves in the vacuum, i.e., the waves propagating towards the positive $z$-direction, from the rest frame to the co-moving frame via the Lorentz transformation. It is expressed as

$$L = \frac{k_0'\gamma}{k_0 q q'}\begin{bmatrix} q^2 - \beta k_0 q_x & \beta k_z q_y \\ -\beta k_z q_y & q^2 - \beta k_0 q_x \end{bmatrix}, \tag{5}$$

where $k_0' = \frac{\omega'}{c}$ and $q' = |\boldsymbol{q}'|$. Similarly, $L^T$ transforms backward propagating waves, and $L^{-1}$ transforms the forward propagating waves from the co-moving frame to the rest frame. Thus, from the viewpoint of the observer in the rest frame, the reflection of electromagnetic waves from the moving slab 2 is calculated by three steps as indicated in $\tilde{R}_2 = LR_2'(L^T)^{-1}$. First, the forward propagating waves are Lorentz transformed from the rest to the co-moving frame by $L$, the reflection is calculated in the co-moving frame by $R_2'$, and the backward propagating waves, represented by $L^T$, are transformed from the co-moving frame to the rest frame by $(L^T)^{-1}$. We note that the inverse of $L$ exists for all the electromagnetic modes that contribute to heat and momentum transfer (see discussions in SI).

Two identities are imposed on the transmission coefficient $\tau_{1\to2}$ in our formalism. Previously, for the case of two anisotropic slabs at rest, direct calculations showed the relation $\tau_{1\to2}(\omega,\boldsymbol{q}) = \tau_{2\to1}(\omega,\boldsymbol{q})$. This relation guarantees that the system satisfies the second law of thermodynamics [47], which requires that $\varphi_{1\to2} = 0$ when $T_1 = T_2$ at $V = 0$. Moreover, the reflection matrix of a reciprocal material satisfies $R_i(\omega,-\boldsymbol{q}) = \sigma_z R_i^T(\omega,\boldsymbol{q})\sigma_z$, where $\sigma_z$ is the Pauli matrix. Therefore, if both slabs are made of reciprocal materials, the transmission coefficient is further constrained as $\tau_{1\to2}(\omega,\boldsymbol{q}) = \tau_{1\to2}(\omega,-\boldsymbol{q})$ [47]. We extend these results for the case of two anisotropic slabs at relative motion. When the slabs are in relative motion, direct calculations show

$$\tau_{1\to2}(\omega,\boldsymbol{q};V) = \tau_{2\to1}(\omega,\boldsymbol{q};V). \tag{6}$$

If the two slabs are made of reciprocal materials, the constraint on the transmission coefficient is derived as

$$\tau_{1\to2}(\omega,-\boldsymbol{q};V)=\tau_{1\to2}(\omega,\boldsymbol{q};-V), \tag{7}$$

where $-V$ on the right-hand side means that object 2 is moving towards the negative $x$-direction. Eq. (7) indicates that the propagation of electromagnetic waves becomes non-reciprocal in the presence of relative motion even when the materials are reciprocal in the respective rest frames. The non-reciprocal propagation induced by moving materials was previously discussed in many different contexts such as the Fizeau drag [50], and the acoustic and optical wave isolation [51,52].

Radiative heat flux and shear stress are not Lorentz scalars. With respect to the Lorentz transformation, the frequency and momentum form a four-vector $q^{\mu}=(\omega/c,q_x,q_y,k_z)$, whereas the electromagnetic fields form a tensor. From Eq. (1), therefore, the radiative heat flux and shear stress also form a four-vector $f^{\mu}=(-\frac{\varphi_{1\to2}}{c},f_{x,1},f_{y,1},f_{z,1})$ as far as the Lorentz boost within the *xy*-plane is concerned. The four-vector transforms between the rest and comoving frame via $f'^{\mu}=\Lambda^{\mu}_{\nu}f^{\nu}$ where $\Lambda^{\mu}_{\nu}$ is the Lorentz boost along the *xy*-plane (see SI for details). Thus, the radiative heat flux and shear stress between the two frames in our system are related as

$$\varphi'_{1\to2}=\gamma\left(\varphi_{1\to2}+f_{x,1}V\right), \tag{8}$$

where $\varphi'_{1\to2}$ is the radiative heat flux from slabs 1 to 2 for the observer in the co-moving frame. We also obtained Eq. (8) by direct calculations.

## 3. HEAT ENGINE

Using the formalism as described in the previous section, we analyze the two-slab structure as schematically shown in Fig. 1(a) from the perspective of using it as a heat engine. This heat engine converts the radiative heat flux into mechanical work driven by the non-equilibrium lateral Casimir force. Practically, a heat engine should be able to self-start. Thus, the two-slab system needs to support a lateral force at $V = 0$ when both slabs are at rest. To fulfill this requirement, two conditions need to be satisfied as can be obtained by examining Eq. (1). First, the two slabs must have different temperatures. For this two-slab system, in general, the radiative heat flux and the shear stress will be identically zero if the slabs are in thermal equilibrium with each other. Second, at least one of the slabs must contain non-reciprocal materials. If both slabs are made entirely of reciprocal materials, the momentum transfer at $\boldsymbol{q}$ and $-\boldsymbol{q}$ cancels out, resulting in null lateral force. This can be seen by observing that the integrand in Eq. (1) for $f_{x,1}$ is an odd function of $q_x$, since the transmission coefficient satisfies the condition $\tau_{1\to2}(\omega, -\boldsymbol{q}; V = 0) = \tau_{1\to2}(\omega, \boldsymbol{q}; V = 0)$ from Eq. (7). These two requirements were also discussed in previous works [36,37].

Based on the discussion above, we consider an $n$-doped Indium Antimonide ($n$-InSb), a well-known magneto-optical material, as the slab material. To break the reciprocity, we externally apply static magnetic fields in the $y$-direction on the two slabs as shown in Fig. 1 (a). The direction of the external magnetic fields is selected to be perpendicular to the intended direction of motion of the slabs along the $x$-direction, so that the degree of asymmetry in the wave propagations will be the greatest and the resulting non-equilibrium lateral force in the $x$-direction will be the largest. In this work, we assume that the external static magnetic fields on slabs 1 and 2 are independently applied in the rest frame and the co-moving frame, respectively. If the external magnetic field is applied on both slabs in the rest frame, the external field on the moving slab in the co-moving frame is composed of static magnetic and electric fields, i.e., $B_y' = \gamma B_y$, and $E_z' = \gamma V B_y$ where $B_y$ is an external static magnetic field in the rest frame, which complicates our theoretical treatment. As long as we consider the heat engine and pump operating at non-relativistic velocities, i.e., $\gamma = 1$ and $V \ll c$ , the two ways of applying the external static magnetic fields give approximately the same results.

Under an external static magnetic field in the positive $y$-direction, the dielectric function of $n$-InSb is given as

$$\frac{\varepsilon(\omega)}{\varepsilon_\infty} = I_3 + \varepsilon_{ph} I_3 - \frac{\omega_p^2}{(\omega + i\gamma)^2 - \omega_c^2} \begin{bmatrix} 1 + i\dfrac{\gamma}{\omega} & 0 & -\dfrac{i\omega_c}{\omega} \\ 0 & \dfrac{(\omega + i\gamma)^2 - \omega_c^2}{\omega(\omega + i\gamma)} & 0 \\ \dfrac{i\omega_c}{\omega} & 0 & 1 + i\dfrac{\gamma}{\omega} \end{bmatrix}, \tag{9}$$

where $I_3$ is the 3 by 3 identity matrix, $\omega_p = \sqrt{n_e e^2/(m_{eff}\varepsilon_0\varepsilon_\infty)}$ is the plasma frequency, $\gamma = 3.39 \times 10^{12}$ rad/s is the electron scattering rate, and $\omega_c = eB/m_{eff}$ is the cyclotron frequency. The carrier concentration and the electron effective mass are taken to be $n_e = 1.07 \times 10^{17}$ cm$^{-3}$, $m_{eff} = 0.022m_e$ where $m_e$ is the bare electron mass, respectively. The bound electron and ions contribution to the dielectric function is incorporated in $\varepsilon_\infty = 15.7$. The phonon contribution $\varepsilon_{ph}$ is given by the Lorenz model as

$$\varepsilon_{ph}(\omega) = \frac{\omega_L^2 - \omega_T^2}{\omega_T^2 - \omega^2 - i\Gamma\omega}, \tag{10}$$

where $\omega_L = 3.62 \times 10^{13}$ rad/s, $\omega_T = 3.39 \times 10^{13}$ rad/s, and $\Gamma = 5.65 \times 10^{11}$ rad/s. All the parameters in the model of *n*-InSb above are taken from [53,54].

Figure 2 shows the performance of the heat engine. We set the temperatures of the slabs to be $T_1 = 305$ K and $T_2' = 300$ K, respectively, and the separation between the slabs to be $d = 10$ nm throughout this study. Note that the small gap between the slabs is not required for the operation as a heat engine, but rather is selected to enhance the power output. In our system, the coupled surface plasmon modes, as well as phonon-polariton modes, have large in-plane momenta as compared with the maximal in-plane momentum of free-space photons at the frequency, i.e., $q_x \gg k_0$, which contributes to larger lateral forces.

Figure 2 (a) shows the radiative heat flux from slabs 1 to 2 as a function of the velocity of relative motion under different magnitudes of the magnetic fields. For two slabs under anti-parallel static magnetic fields, at a fixed velocity, the radiative heat flux decreases as the magnetic field increases in the considered range. The reduction of the radiative heat flux at increasing velocities is due to Doppler shift, and is further related to the interplay of the evolution of surface plasmon and phonon-polariton waves, as well as hyperbolic modes that progressively appear as the magnetic field increases [53]. All these aspects will be discussed in detail later. Practically, it is challenging to apply anti-parallel static magnetic fields for two slabs separated by a nanoscale gap. Therefore, we also consider applying an external static magnetic field to only one of the slabs, i.e., $B_1 = 0$T and $B_2 = -3$T. The results, shown in the green curve in Fig. 2 (a), have qualitatively the same behavior as the cases when anti-parallel magnetic fields are applied to the two slabs. We also note that the shear force is zero at $V = 0$ when the same magnetic fields are applied to both slabs (not shown in Fig. 2). This is consistent with the observation of the symmetry in heat flux spectra when the same magnetic field is applied [47].

Figure 2 (b) shows the shear stress on slab 2. In the absence of the external magnetic field, the shear stress is zero at rest, in consistency with the discussions above. For a moving object, a force that is along the direction of the velocity accelerates the motion of the object. We refer to such a force as a propulsion force. A force with a direction opposite to that of the velocity decelerates the motion of the object. We refer to such a force as a drag force. We also use these wordings of propulsion or drag forces to refer to the different components of a total force. When slab 2 moves in the positive $x$-direction as a result of external mechanical work, the shear stress on slab 2 acts as a drag force, as previously shown for the isotropic materials [48]. This drag force is due mostly to the Doppler shift in the angular frequency. In contrast, in the presence of the external magnetic fields, the shear stress on slab 2 is non-zero even at rest. Therefore, the shear stress here propels slab 2 in the positive $x$-direction, allowing the heat engine to self-start, and acts as a propulsion force after slab 2 starts moving. Figure 2 (b) shows that the shear stress on slab 2 at non-zero $V$ acts as a propulsion force over a wide range of $V$. In this range, within the total shear stress, the propulsion force component is greater than the drag force component. As the velocity of slab 2 increases, the total force acting on slab 2 decreases due to the increasing drag force component and becomes zero at a steady-state velocity. Under the simulation condition, the steady-state velocity is around $V \sim 10^3$ m/s. This steady state is stable: further increase of the velocity results in the net drag force on slab 2. Therefore, to go beyond the steady-state velocity, external work must be applied, and the two-slab structure no longer operates as a heat engine.

We analyze the spectra of the shear stress to reveal the mechanism behind the velocity dependence of the radiative heat flux and shear stress. Figure 3 (a) shows the spectral distribution of the shear stress on slab 2 for different velocities when the two slabs are under the anti-parallel magnetic fields of 3T. First, we discuss the origin of the three peaks and their signed contributions at $V = 0$ m/s. Since the system is translationally invariant and the in-plane momentum is conserved upon reflection, the momentum transfer between slabs can only occur through the emission and absorption processes. Therefore, the momentum transfer to slab 2 can arise either due to the absorption of the thermal emission from slab 1 or the emission of slab 2. Figure 3 (b) shows the transmission coefficient $\tau$ at $V = 0$ m/s as a function of frequency $\omega$ and the in-plane wavevector component along the $x$-direction $q_x$. For other modes with non-zero $q_x$ and $q_y$, the qualitative characteristics of the transmission coefficient are similar (see SI). From Figure 3 (b), surface plasmon and surface phonon-polariton modes at around $\omega = 1.9 \times 10^{13}$ rad/s and $\omega = 3.8 \times 10^{13}$ rad/s, respectively, are supported at $q_x > 0$ but not at $q_x < 0$. When the two slabs are at rest, the number of emitted thermal photons from slab 1 is greater than that from slab 2, $n_B(\omega, T_1) > n_B(\omega, T_2')$. Thus, for these surface modes, the force on slab 2 due to the absorption of thermal radiation emitted by slab 1, which acts as a propulsion force, is greater than the recoil force on slab 2 due to its emission, which acts as a drag force. As a result, the contributions to the lateral force $f_{x,2}(\omega)$ from these two peaks are positive. Similarly, the negative spectral lateral force arises from the surface plasmon modes supported at around $\omega = 4.6 \times 10^{13}$ rad/s at $q_x < 0$.

We note that the lack of symmetry in $\tau(\omega, q_x)$, i.e., the fact that $\tau(\omega, q_x) \neq \tau(\omega, -q_x)$, is essential for the presence of lateral forces. The lateral forces would have been zero if $\tau(\omega, q_x) = \tau(\omega, -q_x)$. Thus, following an argument in Ref. [47], which states that for a reciprocal two-slab system $\tau(\omega, q_x) = \tau(\omega, -q_x)$ at $V = 0$, in reciprocal two-slab systems there is no lateral force at $V = 0$. We also note that the radiative heat flux is dominated by those surface modes and flows from slabs 1 to 2 irrespective of the sign of $q_x$ as shown in Fig. 3 (d).

In Fig. 3 (a), the magnitudes of the three peaks vary with respect to the velocity. Such variations primarily originate from the Doppler shift of the angular frequency, which affects the thermal photon occupation number for the emission from slab 2. The transmission coefficient $\tau$ has a negligible dependency on the velocity range of the heat engine (see SI). Suppose that slab 2 is moving at a non-relativistic velocity $V > 0$, i.e., $\gamma \approx 1$. In the rest frame, consider the thermal radiation emitted by slab 2 with $q_x > 0$ and $\omega$. In the co-moving frame, this thermal radiation has a frequency $\omega' = \omega - q_x V$, which is smaller than $\omega$, and an in-plane momentum $q_x' = q_x - \beta k_0 \approx q_x$, where $\beta = V/c$. Thus, the frequency of the radiation is blue-shifted from the viewpoint of the observer in the rest frame. On the other hand, the number of photons does not change in the two frames and an observer in the rest frame sees the same number of thermally emitted photons as the observer in the co-moving frame, which is given as $n_B(\omega', T_2')$. Thus, the frequency spectrum of the photon number flux that the observer in the rest frame sees is velocity dependent. For the waves with $q_x > 0$, the number of thermal photons at a frequency $\omega$ is exponentially enhanced at $V > 0$ compared to the number at $V = 0$, since $n_B(\omega', T_2') \approx e^{-\frac{\hbar\omega}{k_B T_2'}} e^{\frac{\hbar q_x V}{k_B T_2'}}$ for $\hbar(\omega -$ $q_x V) \gg k_B T_2'$. As a result, as the velocity of slab 2 along the $x$-direction increases, the emission of thermal photons with $q_x > 0$ from slab 2 is exponentially enhanced. This has the effect of enhancing the recoil force, which reduces the total force that propels slab 2. To illustrate this effect, Fig. 3 (c) shows the ratio of the number of net transferred thermal photons per mode at a frequency $\omega$ in the presence of relative motion, $\Delta n_B = n_B(\omega, T_1) - n_B(\omega', T_2')$, to the one without relative motion, $\Delta n_B^0 = n_B(\omega, T_1) - n_B(\omega, T_2')$. The plotted relative velocities are the same as those in Fig. 3 (a) with in-plane wavevectors $q_x = \pm 10^8$ m$^{-1}$ , which has a typical magnitude. For the waves with $q_x > 0$, the net transferred thermal photon number from slabs 1 to 2 is decreased for all the relevant frequencies as the velocity increases. Particularly, the reduction is significant at higher velocities and at lower angular frequencies. This explains the greater reduction of the magnitude of the peak at $\omega = 1.9 \times 10^{13}$ rad/s compared to that at $\omega = 3.8 \times 10^{13}$ rad/s as the velocity increases. For the modes with $q_x < 0$, the observer in the rest frame sees the exponentially suppressed number of thermal photons from slab 2 due to the motion-induced red-shift of the angular frequency. As a result, the net transferred number of thermal photons from slabs 1 to 2 is increased as shown in Fig. 3 (c). As the velocity increases, the drag force component on slab 2 due to the surface modes at around $\omega = 4.6 \times 10^{13}$ rad/s increases. Overall, the increasing velocity of slab 2 results in a greater drag force component on it due to the Doppler shift of the angular frequency. The total force on slab 2 becomes zero at the velocity $V \sim 1.4 \times 10^3$ m/s at which the propulsion force due to the non-reciprocity and the increasing drag force due to the Doppler shift balances.

We numerically observed that the velocity dependence of the radiative heat flux and shear stress is dominated by the exponential change in the thermal occupation number. Then, the expansion of $n_B(\omega', T_2')$ with respect to velocity while setting $V = 0$ in $\tau$ leads to compact expressions of the radiative heat flux and the shear stress. The radiative heat flux and the shear stress can be expressed to the linear order in the non-relativistic limit $\gamma = 1$ as

$$\varphi_{1\to 2}(V) \approx \varphi_{1\to 2}^0 - aV, \qquad f_{x,2}(V) \approx f_{x,2}^0 - bV \tag{11}$$

where $\varphi_{1\to 2}^0 \equiv \varphi_{1\to 2}(V = 0)$ and $f_{x,2}^0 \equiv f_{x,2}(V = 0)$. The coefficients $a$ and $b$ are given as

$$\begin{bmatrix} a \\ b \end{bmatrix} = \int_0^\infty \frac{d\omega}{2\pi} \int \frac{d\boldsymbol{q}}{(2\pi)^2} \hbar \begin{bmatrix} \omega \\ q_x \end{bmatrix} \frac{\hbar q_x}{k_B T_2'} n_B(\omega, T_2')[n_B(\omega, T_2') + 1]\, \tau_{1\to 2}(\omega, \boldsymbol{q}; V = 0). \tag{12}$$

The derivation of Eq. (12) assumes that there is no velocity dependence of the transmission coefficient. The black dashed lines in Fig. 2 (a) show the radiative heat flux calculated by the linear approximation in Eq. (11). The results agree very well with the full calculation up to a velocity $V \sim 5 \times 10^3$ m/s, which is beyond the steady-state velocity. The linear approximation of the shear stress also agrees well with the full calculation as shown in Fig. 2 (b). In Ref. [37], the coefficients $a$ and $b$ are obtained in the operator form by generalizing the linear response expressions of the radiative heat and shear stress to the velocity of relative motion [44] for non-reciprocal systems. Moreover, the coefficient $b$ is derived in the linear polarization basis in the electrostatic limit. Furthermore, the Onsager theorem for fluctuational electrodynamics [44] was used to show that the efficiency is bounded by the Carnot efficiency as far as small perturbations of velocity and temperature difference from the thermodynamic equilibrium are concerned. Here, we provide the linear response expressions in the linear polarization basis and show that they agree well with the full calculations. By taking the electrostatic limit in Eq. (12), the coefficient $b$ should be identical to the friction coefficient in Ref. [37]. Furthermore, the expressions Eq. (11) can be applied to two objects at different temperatures.

The linear expansion of the radiative heat flux and shear stress reveals the competing effects that lead to the maximum thermodynamic efficiency seen in Fig. 2 (c). In non-relativistic velocities, the thermodynamic efficiency of the heat engine is defined for $T_1 > T_2'$ as

$$\eta = \frac{f_{x,2} V}{\varphi_{1\to 2}}, \tag{13}$$

where this efficiency is meaningful only for $f_{x,2} V > 0$. Using Eq. (11), this efficiency can be approximated up to the order of $V^2$ as

$$\eta \approx \frac{\left(f_{x,2}^0 - bV\right)V}{\varphi_{1\to 2}^0 - aV} \approx \frac{f_{x,2}^0}{\varphi_{1\to 2}^0} V - \frac{b}{\varphi_{1\to 2}^0} V^2 + \frac{f_{x,2}^0 a}{(\varphi_{1\to 2}^0)^2} V^2. \tag{14}$$

For small velocities, the efficiency increases linearly with respect to the velocity driven by the propulsion force, $f_{x,2}^{0}$, due to the non-reciprocity of the slab materials. As the velocity increases, the drag force component due to the exponential change in the thermal emission from slab 2, i.e., $bV$, becomes significant, which reduces the efficiency. While the efficiency increases due to the decrease of the heat flux, the overall effects from the terms with the quadratic velocity dependence reduce the efficiency. As a result, the efficiency reaches the maximum at the velocity around

$$V \approx \frac{f_{x,2}^{0}}{2\left(b - \frac{f_{x,2}^{0} a}{\varphi_{1\to 2}^{0}}\right)}. \tag{15}$$

The black dash lines in Fig. 2 (c) show the thermodynamic efficiency calculated by Eq. (14) and show a good agreement with the full calculation. We note that while the linear approximation works very well in our heat engine, the full calculation beyond the linear order and the velocity dependence of the transmission coefficient is critical in the operation of the system as a heat pump at high velocities as we will show in the next section. Finally, we note that the non-zero linear scaling of the radiative heat with respect to the velocity can be considered as a signature of non-reciprocity. When the two slabs are made of reciprocal materials, we can show that, $\varphi_{1\to 2} - \varphi_{1\to 2}^{0} \propto V^2$ (see SI), in contrast to the linear scaling $\propto V$ as seen for the non-reciprocal systems.

Finally, we explore an ideal scenario where the heat engine operates at Carnot efficiency. In our formalism, each frequency is independent. Then, the Carnot efficiency of the heat engine is achieved if and only if the heat engine achieves the Carnot efficiency for each frequency. Since the operating velocity of the heat engine is non-relativistic, we assume $\gamma = 1$. Also, we fix the frequency to be $\omega = \omega_0 > 0$, and consider the monochromatic thermodynamic efficiency $\eta(\omega_0) = \frac{f_{x,2}(\omega_0)V}{\varphi_{1\to 2}(\omega_0)}$ in the following argument.

By rewriting the integral over $q_x$ only in the positive region, the radiative heat flux and the shear stress on slab 2 are expressed as

$$\varphi_{1\to 2}(\omega_0) = \int_0^\infty \frac{dq_x}{2\pi} \hbar\omega_0 \begin{bmatrix} \{n_B(\omega_0, T_1) - n_B(\omega_0 - q_x V, T_2')\}\tau(q_x) \\ +\{n_B(\omega_0, T_1) - n_B(\omega_0 + q_x V, T_2')\}\tau(-q_x) \end{bmatrix}, \tag{16}$$

$$f_{x,2}(\omega_0) = \int_0^\infty \frac{dq_x}{2\pi} \hbar q_x \begin{bmatrix} \{n_B(\omega_0, T_1) - n_B(\omega_0 - q_x V, T_2')\}\tau(q_x) \\ -\{n_B(\omega_0, T_1) - n_B(\omega_0 + q_x V, T_2')\}\tau(-q_x) \end{bmatrix}, \tag{17}$$

where $\tau(q_x) = \int_{-\infty}^{\infty} \frac{dq_y}{2\pi} \tau(q_x, q_y; V)$ and we omit $V$ from the argument to simplify the notation. Here we assume that slab 2 is allowed to move only along the $x$-direction. This assumption appears in the Doppler shift of the angular frequency in the occupation number. The waves with both positive and negative $q_x$ contribute to heat transfer from slabs 1 to 2, but the contribution to the shear stress on slab 2 from modes with negative $q_x$ acts as a drag force and reduces the conversion efficiency. This can be seen in the sign difference in front of the second term in Eqs. (16) and (17), and since $n_B(\omega_0, T_1) > n_B(\omega_0 + q_x V, T_2')$ for $T_1 > T_2'$ in the non-relativistic velocities. Thus, to achieve ideal conversion efficiency from heat to mechanical work, the contributions from modes with negative $q_x$ should be suppressed, i.e., we should set $\tau(-q_x) \to 0$ to obtain the thermodynamic efficiency as

$$\eta(\omega_0) = \frac{\int_0^\infty \frac{dq_x}{2\pi} q_x V [n_B(\omega_0, T_1) - n_B(\omega_0 - q_x V, T_2')]\tau(q_x)}{\int_0^\infty \frac{dq_x}{2\pi} \omega_0 [n_B(\omega_0, T_1) - n_B(\omega_0 - q_x V, T_2')]\tau(q_x)}. \tag{18}$$

To achieve the Carnot efficiency, we assume that for the angular frequency $\omega_0$, only modes with single $q_{x0}$ is supported. In this case, Eq. (18) simplifies to

$$\eta(\omega_0) = \frac{q_{x0} V}{\omega_0}. \tag{19}$$

Thus, the efficiency linearly increases with $V$. The critical velocity $V_C$ where the system reaches steady state can be obtained by setting $n_B(\omega_0, T_1) = n_B(\omega_0 - q_{x0} V, T_2')$, which results in

$$V_c = \frac{\omega_0}{q_{x0}} \left(1 - \frac{T_2'}{T_1}\right). \tag{20}$$

As $V \to V_c$, the heat engine approaches the Carnot efficiency $\eta(\omega_0) \to 1 - T_2'/T_1$. Note, however, that the power output approaches zero as the velocity reaches the critical velocity as expected for a Carnot engine.

In the derivation above we show how to achieve Carnot efficiency from heat transfer at a single frequency $\omega_0$. To achieve the Carnot efficiency over a broad range of frequencies, the critical velocity $V_c$ as determined in Eq. (20) must be independent of $\omega_0$. Thus, the required dispersion is

$$\omega = \frac{V_C}{1 - \frac{T_2'}{T_1}} q_x, \qquad \text{with } q_x > 0 \tag{21}$$

and no waves in the opposite direction $q_x < 0$.

## 4. HEAT PUMP

In the operation of the heat engine, the velocity of relative motion eventually reaches a steady state at which the output work is zero. In this section, we consider the situation where external work is applied on slab 2 to further increase the velocity beyond the steady state velocity. In this case, the external work provides energy input to the system by operating against the drag force. As the velocity increases, the thermal emission from slab 2 is enhanced, which results in enhanced energy transfer from slabs 2 to 1. At sufficiently high velocity, we show that the radiative heat flows from slab 2 at a lower proper temperature to slab 1 at a higher proper temperature. Hence, the system operates as a photonic heat pump that utilizes external mechanical work to pump from a lower to a higher temperature object.

We consider the system under external work as shown in Fig. 1 (b). We first provide an intuitive explanation of the mechanism of the photonic heat pump by considering an electromagnetic mode with $\omega$ and $\boldsymbol{q} = (q_x, 0)$. When slab 2 is moving at the relative velocity $V$ in the positive $x$-direction, the net radiative heat flux from slabs 1 to 2 due to the modes with $q_x > 0$ satisfies $\varphi_{1\to2}(q_x) \propto \hbar\omega[n_B(\omega, T_1) - n_B(\gamma(\omega - q_x V), T_2')]\tau(q_x; V)$. As we discussed in the previous section, the increasing velocity of slab 2 induces the exponential enhancement of the thermal emission due to this mode from slab 2 for the observer in the rest frame. When the velocity is sufficiently high, the net radiative heat flux from these modes flows from slabs 2 to 1 even when $T_1 > T_2'$ and hence $\varphi_{1\to2}(q_x) < 0$. The electromagnetic modes that contribute to cooling must have $n_B(\omega, T_1) < n_B(\gamma(\omega - q_x V), T_2')$ , and thus must have the dispersion relation satisfying the constraint:

$$\omega < \frac{V}{1 - \frac{T_2'}{\gamma T_1}} q_x \approx c \frac{T_1}{T_1 - T_2'} \frac{V}{c} q_x \, . \tag{22}$$

For non-relativistic velocities, the constraint in Eq. (22) implies that $\omega \ll c q_x$ and thus requires a dispersion relation that lies outside the light cone. Therefore, the near-field radiative heat transfer is critical in order to achieve the heat pump.

For the modes with $q_x < 0$, the radiative heat flux is $\varphi_{1\to2}(-|q_x|) \propto \hbar\omega[n_B(\omega, T_1) - n_B(\gamma(\omega + |q_x|V), T_2')]\tau(-|q_x|; V)$. When $T_1 > T_2'$, these modes contribute to heating only. Therefore, to achieve cooling after integrating contributions from all modes, it is also required that the transmission coefficient exhibits strong enough non-reciprocity $\tau(q_x; V) \gg \tau(-q_x; V)$.

Based on the discussion above, two conditions are required to achieve cooling at non-relativistic velocities. First, two slabs must support the electromagnetic modes whose dispersions satisfy the constraints of Eq. (22). This can be achieved by the coupled surface modes in the near-field. Second, considering that the total radiative heat flux is integral over the frequency and wavevector spaces, the surface modes that satisfy the dispersion Eq. (22), which contribute to cooling, must dominate over the other modes that contribute to heating. This is achieved by the non-reciprocity that allows strong directional surface modes.

Figure 4 (a) and (b) show the radiative heat flux from slabs 1 to 2 and the shear stress on slab 2, respectively, for the same materials and temperatures as the heat engine in Fig. 2, but in a range of higher velocity. The shear stress on slab 2 is negative and external work needs to be applied in order to move slab 2 at such velocities. In the absence of the external magnetic field, the radiative heat flux becomes negative for the velocity greater than $10^5$ m/s, where the heat flows from slab 2 at a lower proper temperature to slab 1 at a higher proper temperature. This shows that structures made of reciprocal material can operate as a heat pump due to the non-reciprocal wave propagations as induced by the Doppler effect. With reciprocal materials, the heat pump cannot occur at non-relativistic velocities if the velocity dependence of the transmission coefficient is ignored because the heating from the modes with $q_x > 0$ always overcome the cooling from the modes with $q_x < 0$. Hence, the velocity dependence of the transmission coefficient is essential in the heat pump regime when reciprocal materials are used. The application of the external static magnetic fields can lower the velocity at which the cooling occurs; with the presence of anti-parallel magnetic fields of 6 T, the onset of the cooling occurs at a velocity of $V \approx 2 \times 10^4$ m/s, which is almost an order of magnitude lower as compared with that without the magnetic fields. Comparing the cases of $B_1 = B_2 = 0$ T and $B_1 = -B_2 = 3$ T, the greater amount of radiative heat can be pumped by applying less amount of external mechanical work when the velocity is below $V = 4.5 \times 10^5$ m/s. This shows that the non-reciprocal materials can enhance the performance of the heat pump. The magnitude of pumped radiative heat further increases under $B_1 = -B_2 = 6$ T while the shear stress exhibits a moderate change compared to the cases of $B_1 = B_2 = 0$ T and $B_1 = -B_2 = 3$ T. As a result, the heat pump performance further increases with increasing magnetic fields. Figure 4 (c) shows the coefficient of performance (COP) defined as $\mathrm{COP} = \frac{\varphi_{1\to 2}}{f_{x,2}V}$ for $T_1 > T_2'$. The onset of the heat pump operation at lower velocities and the enhancement of the performance by the external magnetic fields can be clearly seen.

Figure 5 shows the spectral radiative heat flux for different velocities under the application of the anti-parallel external magnetic fields of 3T. To understand the evolution of the spectral radiative heat flux as the velocity increases, Fig. 6 (a) and (b) show the transmission coefficients and, (c) and (d) show the spectral radiative heat flux for the modes with $\boldsymbol{q} = (q_x, 0)$ and for the two different velocities in the case of $B_1 = -B_2 = 3$ T. For other modes with non-zero $q_y$, the qualitative characteristics of the transmission coefficient are similar (see SI). As discussed in Section II, the increase of the velocity reduces the net radiative heat flux from slabs 1 to 2 due to the exponential change of the emission from slab 2. For the waves with dispersions that satisfy the constraint in Eq. (22), the exponential enhancement of thermal emission from slab 2 exceeds the thermal emission from slab 1, resulting in the net negative radiative heat flux and contributing to cooling. In Fig. 6 (c) and (d), the right-hand side of Eq. (22) is plotted as the “cooling line”. Any mode that lies below the cooling line contributes to cooling. As shown in Fig. 6 (c), at $V = 10^4$ m/s, the large part of the coupled surface plasmon modes around $\omega = 1.9 \times 10^{13}$ rad/s, which has $q_x > 0$ and $q_y = 0$, lies below the cooling line. Moreover, there are no surface modes near this frequency with $q_x < 0$ due to non-reciprocity. Thus, there is a significant contribution to cooling from these modes at $q_y = 0$ . For the other waves with non-zero $q_y$, the net contribution to cooling is smaller due to the smaller portion of the surface waves that contribute to cooling. But the net cooling contribution persists after integrating over all $q_y$. As a result, the radiative heat flux at around $\omega = 1.9 \times 10^{13}$ rad/s is net negative as shown in Fig. 5. For the coupled surface phonon polariton modes at around $\omega = 3.8 \times 10^{13}$ rad/s, the cooling line intersects with its dispersion relation. Thus, the magnitudes of their contributions to heating and cooling are similar, resulting in the small radiative heat flux at $V = 10^4$ m/s as shown in Fig. 5. The coupled surface plasmon modes at around $\omega = 4.6 \times 10^{13}$ rad/s mostly have $q_x < 0$. Thus, the net radiative heat flux from slabs 1 to 2 is enhanced due to the exponential suppression of thermal emission from slab 2. As a result, the spectral radiative heat flux increases at $V = 10^4$ m/s compared to that at rest. At $V = 10^4$ m/s, while the spectral radiative heat flux at the lower frequencies contributes to cooling, the overall radiative heat flux still flows from slabs 1 to 2. However, the magnitude of radiative heat flux from slabs 1 to 2 at $V = 10^4$ m/s is suppressed to almost half of that at rest.

As the velocity of relative motion further increases to $V = 1.5 \times 10^5$ m/s at which the COP reaches the maximum, the magnitude of the negative radiative heat flux from the modes below the cooling line further increases as shown in Fig. 6 (d). At this velocity, the negative radiative heat flux dominates over the positive radiative heat flux even after integrating over the frequency and wavevector space, which results in the heat pump as discussed in Fig. 4 (a). Furthermore, Fig. 6 (d) shows that the dominant contribution to cooling arises from the coupling between the surface plasmon and surface phonon polariton modes at around $\omega = 3.8 \times 10^{13}$ rad/s and $q_x = 2 \times 10^8$ m$^{-1}$ due to the Doppler shift of the surface plasmon modes, which will be discussed in detail below. The contribution from this coupled mode appears as a strong peak in Fig. 5.

Another observation from Fig. 6 is the strong modification of the dispersion relation of the surface modes compared with the case of $V = 0$ shown in Fig. 3 (b) and (d). In Fig. 6 (d), the dispersion of the surface waves for $q_x > 0$ and $q_x < 0$ shows upward and downward tilting, respectively, due mostly to the Doppler shift of the angular frequency. Suppose that the resonance condition of a surface wave on slab 1 is satisfied in the rest frame, i.e., for a given $q_x$, $\omega = f(q_x)$, where $f$ is the dispersion of the surface wave. In the co-moving frame, the same surface mode is supported on the surface of slab 2, i.e., $\omega' = f(q_x')$. When observing this surface wave from the rest frame, the Lorentz transformation of the frequency and wavevector results in the surface wave supported at the interface of the moving slab as $\omega \approx q_x V + f(q_x)$. Thus, the dispersion curve shows upward or downward tilting by $q_x V$ depending on the sign of $q_x$. The Lorentz transformation of the frequency is the effects of Doppler shift. This Doppler shift of the angular frequency results in the off-resonance of the two surface waves that are in resonance at rest. The decoupling of the surface waves typically causes a smaller transmission coefficient at the original resonant frequency and instead the transmission coefficient spreads over a wider frequency range around the original resonant frequency. Interestingly, the opposite can happen at high velocities. At high velocities, two surface modes supported on two slabs that are off-resonance in the absence of the relative motion can be brought into resonance in the presence of the relative motion. To show this clearly, the dispersion of the surface waves supported between the two slabs at relative motion with non-relativistic velocities in the Voigt configuration is derived as

$$\left[-\frac{\omega'}{\omega}+\frac{\varepsilon_d}{\varepsilon_V^{(1)}\varepsilon_V^{(2)}}\left(\frac{\varepsilon_{xz}^{(2)}}{\varepsilon_{xx}^{(2)}}\frac{q_x}{k_z}+i\frac{\kappa_2}{k_z}\right)\left(\frac{\varepsilon_{xz}^{(1)}}{\varepsilon_{xx}^{(1)}}\frac{q_x}{k_z}-i\frac{\kappa_1}{k_z}\right)\right]\tanh(-ik_zd) \\ -\frac{\varepsilon_d}{\varepsilon_V^{(1)}}\left(\frac{\varepsilon_{xz}^{(2)}}{\varepsilon_{xx}^{(2)}}\frac{q_x}{k_z}+i\frac{\kappa_2}{k_z}\right)+\frac{\omega'}{\omega}\left(\frac{\varepsilon_{xz}^{(1)}}{\varepsilon_{xx}^{(1)}}\frac{q_x}{k_z}-i\frac{\kappa_1}{k_z}\right)=0, \tag{23}$$

where $\varepsilon_d = 1$ is the dielectric function of the vacuum gap, $\kappa_1 = \sqrt{q_x^2 - \varepsilon_V^{(1)} k_0^2}$ , $\kappa_2 = \sqrt{q_x^2 - \varepsilon_V^{(2)} k_0^2}$, and $\varepsilon_V^{(i)} = \varepsilon_{zz}^{(i)} + \left(\varepsilon_{xz}^{(i)}\right)^2/\varepsilon_{zz}^{(i)}$ $i = 1,2$ where the subscripts stand for the elements of the dielectric function tensor. For slab 2, the dielectric functions are evaluated in the co-moving frame, i.e., $\varepsilon_{zz}^{(2)} = \varepsilon_{zz}^{(2)}(\omega')$, $\varepsilon_{xz}^{(2)} = \varepsilon_{xz}^{(2)}(\omega')$. The solutions of Eq. (23) in the lossless limit are plotted in Fig. 6 by the white dotted lines for the region where $\omega' > 0$. The dispersion relation Eq. (23) reproduces the velocity dependence of the surface waves by the full calculations very well. In Fig. 6 (d), the dispersion curve Eq. (23) clearly shows that the surface plasmon modes that are blue-shifted due to the Doppler effect come into resonance with the surface phonon polariton modes at around $\omega = 3.8 \times 10^{13}$ rad/s. Although the transmission coefficient as a result of the resonance is moderate, i.e., $\tau \sim 0.15$, the modes are supported at the high in-plane momentum. As a result, the contribution to the radiative heat flux becomes significant as we discussed in Fig. 5. Note however that such modes are supported at the high in-plane momentum, the incorporation of non-local effects into our dielectric function models can suppress the resonance. Nevertheless, this result shows that the Doppler shift can provide an additional control knob to engineer two off-resonant modes into resonance. We also note that the photonic heat pump can occur without this coincidental resonance. In fact, the heat pump can be achieved by magnetized plasma by turning off the contribution from the phonon polaritons, i.e., $\varepsilon_{ph} = 0$ (see SI).

Finally, we emphasize that the heat pump does not require electromagnetic modes with negative frequencies in the co-moving frame, i.e., $\omega' < 0$. In such regimes, from the viewpoint of the observer in the rest frame, the moving slab behaves as a gain material. In general, the modes with $\omega' < 0$ contribute to the cooling since $n_B(\omega', T_2')\tau_{1\to2} > 0$. The black dashed line in Fig. 6 (d) sets the boundary below which the frequency in the co-moving frame becomes negative. One can see that all relevant electromagnetic modes contributing to the heat pump exist in the positive frequency domain. However, as the relative velocity increases further the significant contribution comes from the modes with negative frequencies in the co-moving frame. The contribution of the modes with $\omega' < 0$ to the radiative heat flux under the anti-parallel magnetic fields of 3T is shown in Fig 4 (a). At $V = 1.5 \times 10^5$ m/s where the maximum COP is achieved, the contribution from the negative frequency modes is less than 10% of the total radiative heat flux and even less at lower velocities. Thus, those modes are not required to achieve the heat pump. As the velocity increases, the contribution to the radiative heat flux from the modes with $\omega' < 0$ becomes dominant.

## 5. RELATIVISTIC THERMODYNAMIC EFFICIENCY

The thermodynamic efficiency of the heat engine considered in this work is defined as Eq. (13). For situations where the velocity of relative motion is non-relativistic, i.e., $\gamma \approx 1$, we showed that such definition of the thermodynamic efficiency of the heat engine approaches the Carnot efficiency $1 - \frac{T_2'}{T_1}$ when the electromagnetic modes satisfy the dispersion in Eq. (21). In principle, we can conceive of a heat engine where two slabs are in relative motion at relativistic velocities. For such systems, as we will show below, the efficiency as defined in Eq. (13) is problematic. For the observer in the rest frame, this definition leads to the efficiency limit of $1 - \frac{T_2'}{\gamma T_1}$. On the other hand, for the observer in the co-moving frame, it leads to the limit of $1 - \frac{\gamma T_2'}{T_1}$. Therefore, this definition results in the thermodynamic limit that depends on the frame of reference. This indicates that in some inertial frame, the efficiency limit is higher than what one may define as the sensible definition of the Carnot efficiency $1-\frac{T_2'}{T_1}$. Moreover, in the limit of $V \to c$, it indicates that the perfect conversion of heat into work is possible in the rest frame since $\gamma \to \infty$.

The difficulty of the efficiency definition in Eq. (13) originates from the definition of $f_{x,2}V$ as the useful work in our system, which does not take the relativity effects into account. It was pointed out that with different choices of the forms of useful work, the relativistic Carnot efficiency in the rest frame can take different forms $\eta = 1-\frac{\gamma^a T_2'}{T_1}$ where $a = 0, \pm 1$ [55]. Thus, the problem is also closely related to the Lorentz transformation of the temperature of the moving material $T_2 = T_2'\gamma^a$ where $T_2$ is the temperature of the moving material in the rest frame, which has been a topic of discussion for decades and remains to be answered [56].

In this section, our purpose is to derive the expression of relativistic thermodynamic efficiency for our system that considers the relativistic effects. We focus on the operation of the system as a heat engine to derive the relativistic thermodynamic efficiency, but a similar argument should follow for the operation as a heat pump. First, we show that the efficiency bound based on the non-relativistic thermodynamic expression Eq. (13) depends on the frame of reference in the relativistic regime. Adopting the viewpoint of Landsberg [55] that the Carnot limit should not depend on the choice of a particular reference frame, we discuss two contributions as non-extractable work by using the relativistic thermodynamics [57]. By considering these contributions to the definition of useful work, we derive the relativistic thermodynamic efficiency and show that it is bounded by the Carnot efficiency which is independent of the frame of reference.

First, we can prove the following identity (see detailed derivation in SI)

$$T_1 f_{x,2} V \leq \left| T_1 - \frac{T_2'}{\gamma} \right| \operatorname{sgn}\left( T_1 - \frac{T_2'}{\gamma} \right) \varphi_{1\to 2}. \tag{24}$$

By applying Eq. (24) to the expressions of the conventional thermodynamic efficiency of Eq. (13), we can find the efficiency bound as

$$\eta = \begin{cases} \dfrac{f_{x,2}V}{\hat{\varphi}_{1\to 2}} \leq 1 - \dfrac{T_2'}{\gamma T_1} & \text{if } T_1 > \dfrac{T_2'}{\gamma} \\ \dfrac{f_{x,2}V}{\hat{\varphi}_{1\to 2} + f_{x,2}V} \leq 1 - \dfrac{\gamma T_1}{T_2'} & \text{if } T_1 < \dfrac{T_2'}{\gamma} \end{cases}, \tag{25}$$

where $\hat{\varphi}_{1\to2} = \text{sgn}\left(T_1 - \frac{T_2'}{\gamma}\right)\varphi_{1\to2}$ and this definition is meaningful only if $f_{x,2}V \geq 0$. Thus, the conventional thermodynamic efficiency expression in Eq. (25) leads to the bound that depends on the frame of reference due to the dependence on the Lorentz factor $\gamma$.

The contribution that cannot be extracted as useful work originates from the change of momentum of slab 2 as a result of heat transfer. In relativistic mechanics, the changes of energy and momentum are directly related. In the rest frame, the *x*-component of the momentum of a moving body at the velocity $V$ in the *x*-direction is related to its energy $E$ by $G_x = \frac{E}{c^2}V$. To identify the part of work that is non-extractable, we separate slab 2 into a hypothetical engine connected to an infinitely large heat sink at $T_2'$. Both the hypothetical engine as well as the heat sink move at the velocity $V$ as shown in Fig. 7 (a). We consider the control volume with the surface area $dA = dxdy$ on the *xy*-plane that encloses the engine. For this control volume, we consider the first law of relativistic thermodynamics from the viewpoint of an observer in the rest frame [57]

$$dE = \delta Q + \delta W + VdG_x. \tag{26}$$

This relativistic first law of thermodynamics dictates that the energy change $dE$ of the control volume is due to the heat input to the control volume, $\delta Q$, the work done to the control volume, $\delta W$, and the change of the momentum associated with the heat transferred to the control volume, $VdG_x$. We apply Eq. (26) to one cycle of the heat engine and we denote the period as $\Delta t$. The heat transferred to the control volume is $\frac{\delta Q}{A\Delta t} = \varphi_{1\to2}$. The work done to the control volume is $\frac{\delta W}{A\Delta t} = -f_{x,2}V$. The contribution to the energy change due to the change of the momentum is $\frac{VdG_x}{A\Delta t} = \beta^2\frac{dE}{A\Delta t}$. In the operation of a heat engine, the system must return to the same thermodynamic state as the initial state after one cycle. In relativistic thermodynamics, it means that the system must retain not only the same energy but also the same momentum after one cycle. Thus, in order to keep the momentum of the moving medium unchanged after one cycle, the amount $\beta^2\frac{dE}{A\Delta t}$ has to be rejected to the heat sink and cannot be used as useful work. The energy change of the control volume after one cycle is given from Eq. (26) as

$$\frac{dE}{A\Delta t} = \gamma^2\left(\varphi_{1\to2} - f_{x,2}V\right) = \gamma\varphi'_{1\to2}. \tag{27}$$

Subtracting the non-extractable work $\beta^2\frac{dE}{A\Delta t}$ from $f_{x,2}V$, we have the net useful work from the engine as

$$W_1 = f_{x,2}V - \gamma^2\beta^2\left(\varphi_{1\to2} - f_{x,2}V\right). \tag{28}$$

Another way to interpret Eq. (28) is that the total energy change in the control volume after one cycle must be zero. Thus, the amount of energy given in Eq. (27) must flow into the heat sink to keep the momentum unchanged. Then, the useful work is given as $W_1 = \varphi_{1\to2} - \gamma\varphi'_{1\to2}$, which is the same result as Eq. (28).

The other missing contribution in Eq. (25) is the additional work that can be extracted as a result of the energy increase of slab 2 after one cycle. Consider the operation of the heat engine from the state where the two slabs are at rest. To operate the heat engine at finite velocity $V$, we need work to accelerate slab 2. In our heat engine, this work is provided by the propulsion force driven by the non-equilibrium lateral Casimir force. After the operation of the heat engine at the velocity $V$, we decelerate slab 2 until the two slabs are at rest. Under the adiabatic approximation for the acceleration and deceleration, the amount of work required for the two processes is given as the differences in the kinetic energies. In non-relativistic physics, the work required for the two processes are the same because the mass of slab 2 is invariant before and after the operation of the heat engine. However, in the relativistic cases, the work done by slab 2 in the deceleration is greater than the work done to slab 2 to accelerate due to the energy increase of slab 2 after the acceleration process. In principle, this difference between the two can be extracted as useful work.

To estimate the portion of the work, we consider operating the heat engine from the slabs at rest and determine the amount of work necessary to accelerate slab 2 to the velocity at which we operate the heat engine and decelerate it back to rest after the operation of the heat engine as shown in Fig. 7 (b). We consider the adiabatic acceleration and deceleration of slab 2 from the point of the observer in the rest frame. Then, the work required to accelerate slab 2 to the velocity $V$ is given by the energy difference $W_a = (\gamma - 1)E$ where $E$ is the energy of slab 2 at rest. During the operation of the heat engine over the time period $\Delta t$, the increase of the energy of slab 2 evaluated in the co-moving frame is $\varphi'_{1\to2}A'\Delta t' = \varphi'_{1\to2}A\Delta t$ where we used $A\Delta t = A'\Delta t'$, which results from the fact that the number of photons is Lorentz invariant. Thus, the work that can be extracted by decelerating slab 2 to rest is given as $W_d = (\gamma - 1)(E + \varphi'_{1\to2}A\Delta t)$. This is greater than the work required for the acceleration process. Then, the net mechanical work that can be extracted is given as the difference between them as $W_d - W_a = (\gamma - 1)\varphi'_{1\to2}A\Delta t = \gamma(\gamma - 1)\left(\varphi_{1\to2} - f_{x,2}V\right)A\Delta t$ where we used Eq. (8). Thus, the portion of the work per unit area per cycle contributed by the mechanical energy of slab 2 is given as

$$W_2 = \gamma(\gamma - 1)\left(\varphi_{1\to2} - f_{x,2}V\right), \tag{29}$$

which needs to be added to the useful work.

Overall, the amount of useful work per unit area per second is

$$W_{net} = W_1 + W_2 = f_{x,2}V - \gamma^2\beta^2\left(\varphi_{1\to2} - f_{x,2}V\right) + \gamma(\gamma - 1)\left(\varphi_{1\to2} - f_{x,2}V\right). \tag{30}$$

And the relativistic thermodynamic efficiency for $T_1 > \frac{T'_2}{\gamma}$ is given as

$$\eta_{rel} = \frac{W_{net}}{\varphi_{1\to2}} = \frac{\gamma f_{x,2}V - (\gamma - 1)\varphi_{1\to2}}{\varphi_{1\to2}}. \tag{31}$$

In the non-relativistic limit, Eq. (31) goes back to the conventional nonrelativistic thermodynamic efficiency. By using the efficiency bound Eq. (25), we find that the relativistic thermodynamic efficiency is bounded by the Carnot efficiency that is independent of the frame of reference

$$\eta_{rel} \leq 1 - \frac{T'_2}{T_1}. \tag{32}$$

The appreciable difference between the relativistic thermodynamic efficiency and the non-relativistic one is observed when the velocity of relative motion is a sizable fraction of the speed of light.

## 6. CONCLUSIONS

In summary, we showed that a system consisting of two semi-infinite parallel slabs with different temperatures can work as a photonic heat engine driven by non-equilibrium lateral Casimir forces. To realize it, one of the materials must break Lorentz reciprocity. Also, we demonstrated that in this system a sufficiently high velocity of relative motion realized by external work can pump radiative heat from the low to high temperature objects, thus this system can also operate as a photonic heat pump. Non-reciprocal wave propagations induced by the relative motion can realize the heat pump and the use of non-reciprocal materials can further reduce the amplitude of the required velocity of relative motion to achieve the heat pump. We proved that the relativistic thermodynamic efficiency of the photonic heat engine and pump is bounded by the Carnot efficiency and revealed the ideal dispersion of materials that approaches the limit. Our results point to a way of thermal energy harvesting and cooling by non-equilibrium Casimir forces enabled by breaking Lorentz reciprocity.

## Acknowledgment

This work is supported by a MURI program from the U. S. Army Research Office (Grant No. W911NF-19-1-0279). We thank Dr. Mohammed Benzaouia for a fruitful discussion.

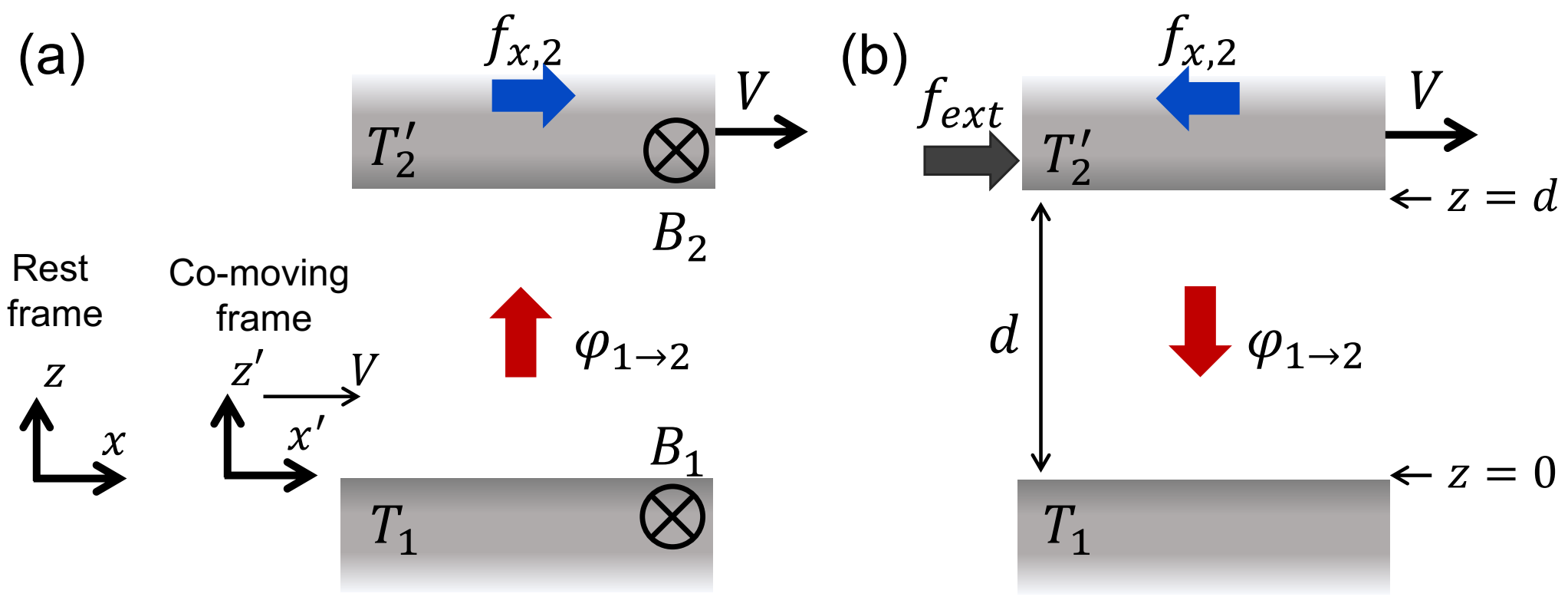

FIG.1. Two semi-infinite parallel slabs separated by a vacuum gap $d$ that are moving relative to each other at constant velocity $V$ along the $x$-direction. Two slabs are at the proper temperatures $T_1$ and $T_2'$ in the rest and co-moving frames, respectively. The physical quantities with the prime mark are defined in the co-moving frame. (a): the system operating as a heat engine where the net heat flux $\varphi_{1\to2}$ from slabs 1 to 2 is converted into work driven by non-equilibrium lateral Casimir force $f_{x,2}$ on slab 2 under proper directions of external magnetic fields. (b): the system operating as a heat pump where the external work done by the force $f_{ext}$ pumps heat from slab 2 at a low proper temperature to slab 1 at a higher proper temperature. External magnetic fields are not necessary.

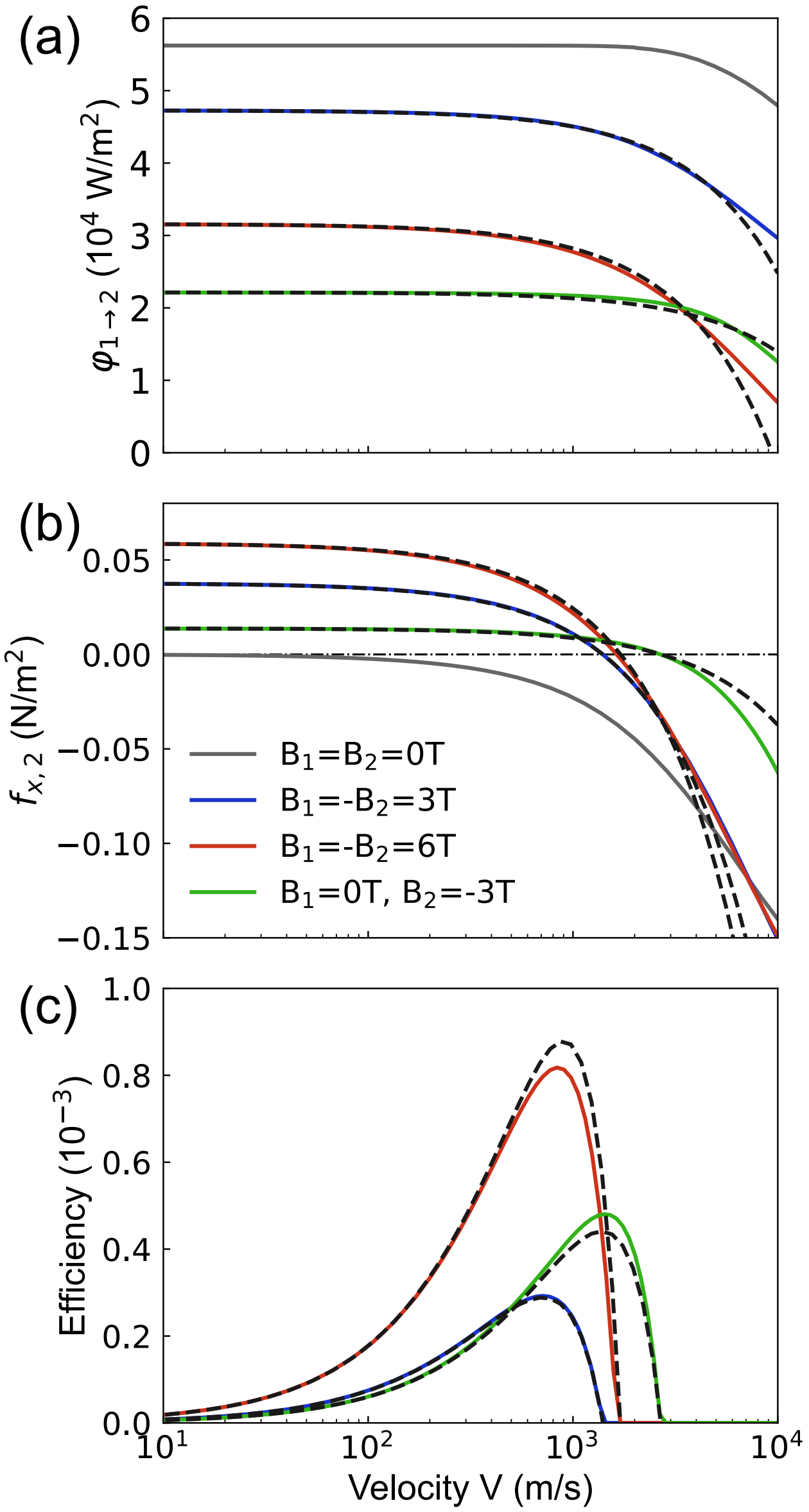

FIG. 2. (a) Radiative heat flux from slabs 1 to 2, (b) shear stress on slab 2, and (c) thermodynamic efficiency of the heat engine as a function of the velocity of relative motion. Two slabs made of $n$-InSb are at $T_1 = 305$ K and $T_2' = 300$ K and separated by $d = 10$ nm. The results for different magnitudes of external magnetic fields are shown. The dash-dot line in panel (b) describes $f_{x,2} = 0$. The results of the linear approximation Eq. (11) and Eq. (14) are plotted in the dashed lines.

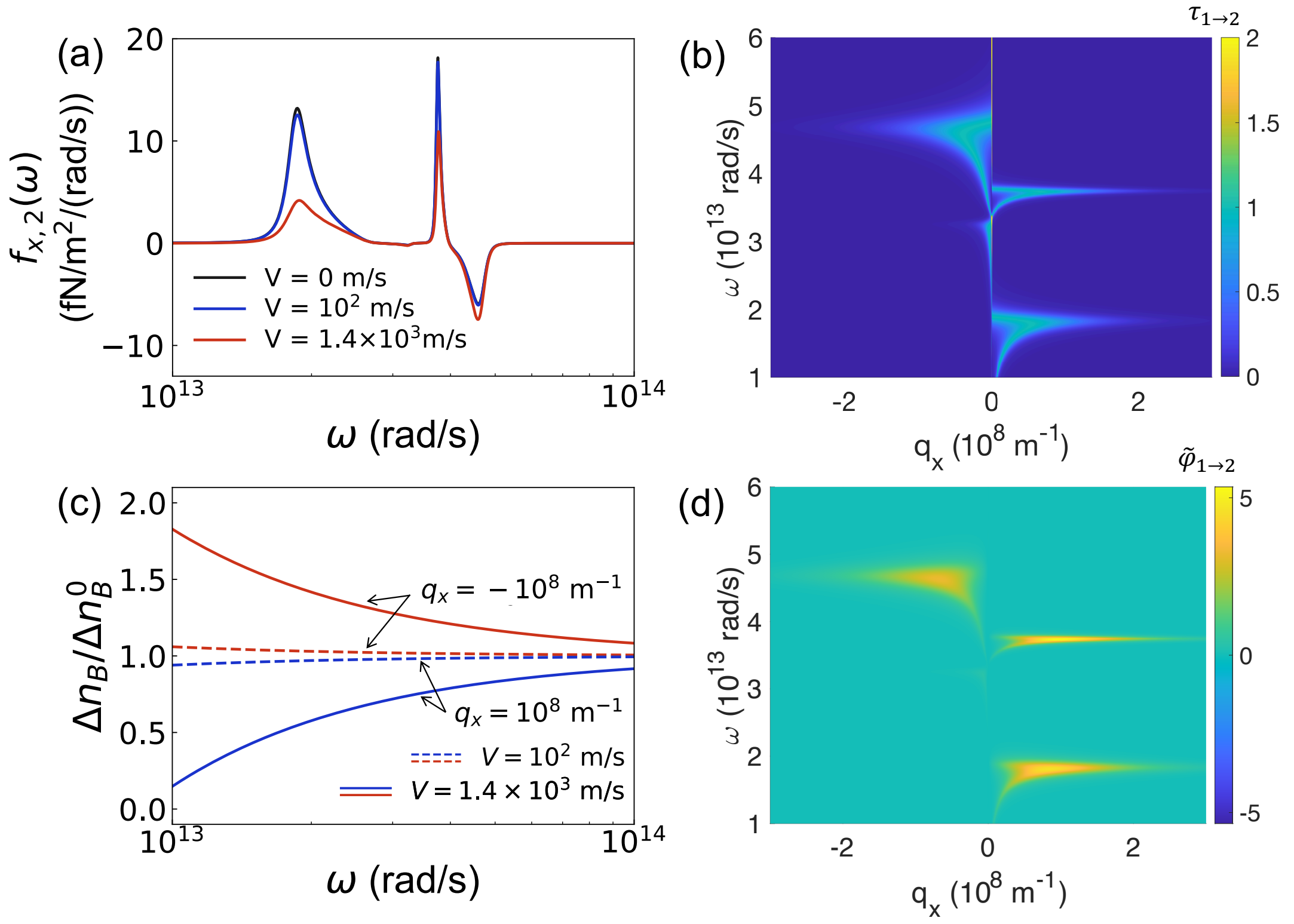

FIG. 3. (a) Spectral shear stress on slab 2 for different velocities of relative motion. (b) Plot of the transmission coefficient $\tau$ as a function of the frequency $\omega$ and $x$-component of the wavevector $q_x$ when two slabs are at rest. The transmission coefficient is plotted for the modes with in-plane wavevectors along the $x$-direction, i.e., $\boldsymbol{q} = (q_x, 0)$. The magnetic fields are $B_1 = -B_2 = 3\text{T}$. (c) The ratio of the net transferred thermal photon number with relative motion $\Delta n_B = n_B(\omega, T_1) - n_B(\omega', T_2')$ to the one without relative motion $\Delta n_B = n_B(\omega, T_1) - n_B(\omega, T_2')$ for fixed $q_x = \pm 10^8$ m$^{-1}$ at $V = 10^2$ m/s and $V = 1.4 \times 10^3$ m/s. (d) Plot of mode-resolved radiative heat flux $\tilde{\varphi}_{1\to2}(\omega, q_x, q_y)$ defined as $\varphi_{1\to2} = \int_0^\infty d\omega \int_{-\infty}^{\infty} d\boldsymbol{q}\, \tilde{\varphi}_{1\to2}(\omega, q_x, q_y)$ for the same condition as panel (b). The unit is $[10^{-15} \frac{\text{W}\cdot\text{s}}{\text{rad}}]$. The other physical parameters are the same as those in Fig. 2.

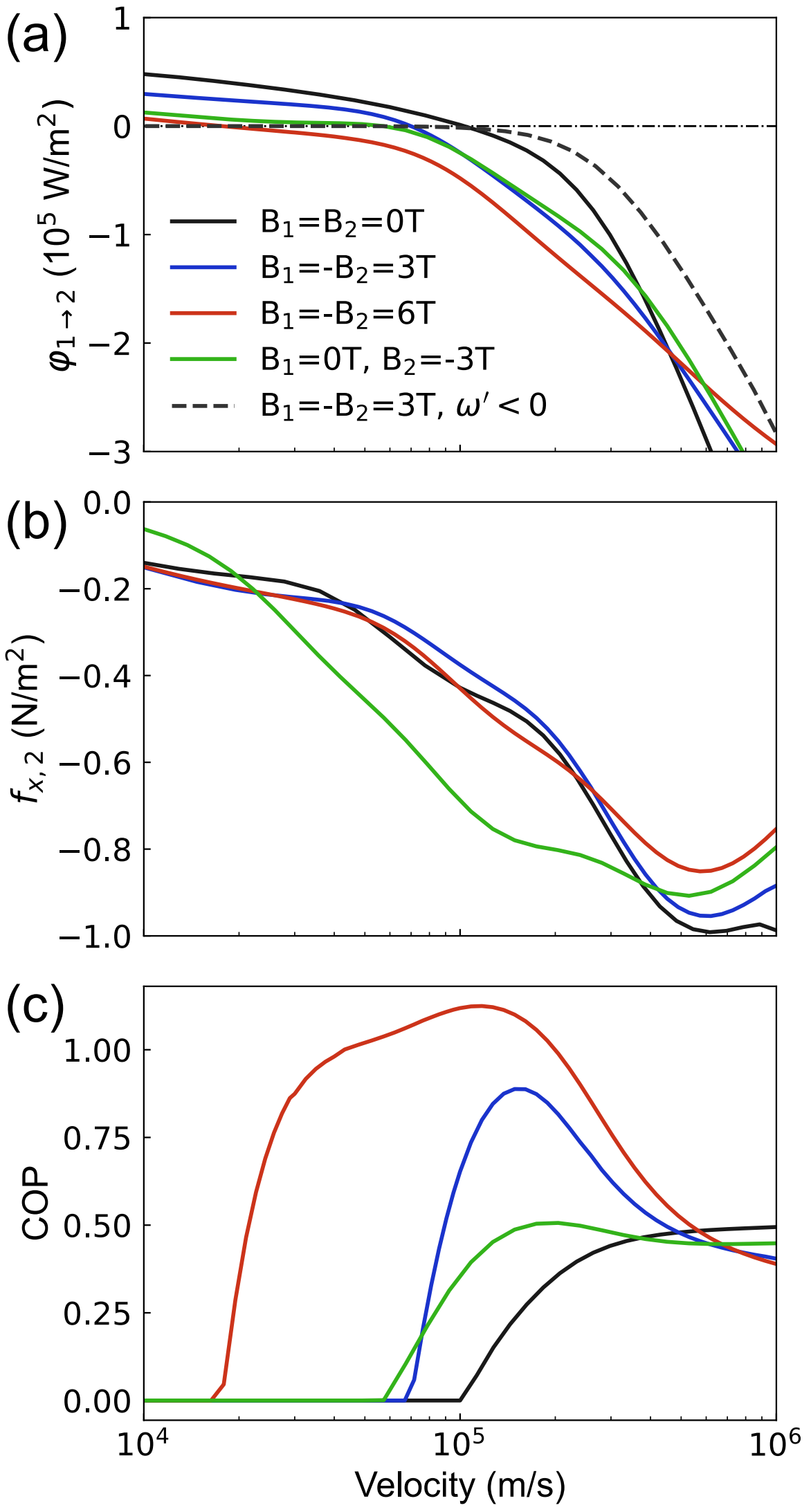

FIG. 4. (a) Radiative heat flux from slabs 1 to 2, (b) shear stress on slab 2, and (c) coefficient of performance of the heat pump as a function of the velocity of relative motion. The physical parameters are the same as those in Fig. 2. The dash-dot line in panel (a) describes $\varphi_{1\to2} = 0$. The COP is only defined for the region of $\varphi_{1\to2} < 0$ where the system operates as the heat pump. The black dash line in panel (a) represents the radiative heat flux from the modes with $\omega' < 0$ only.

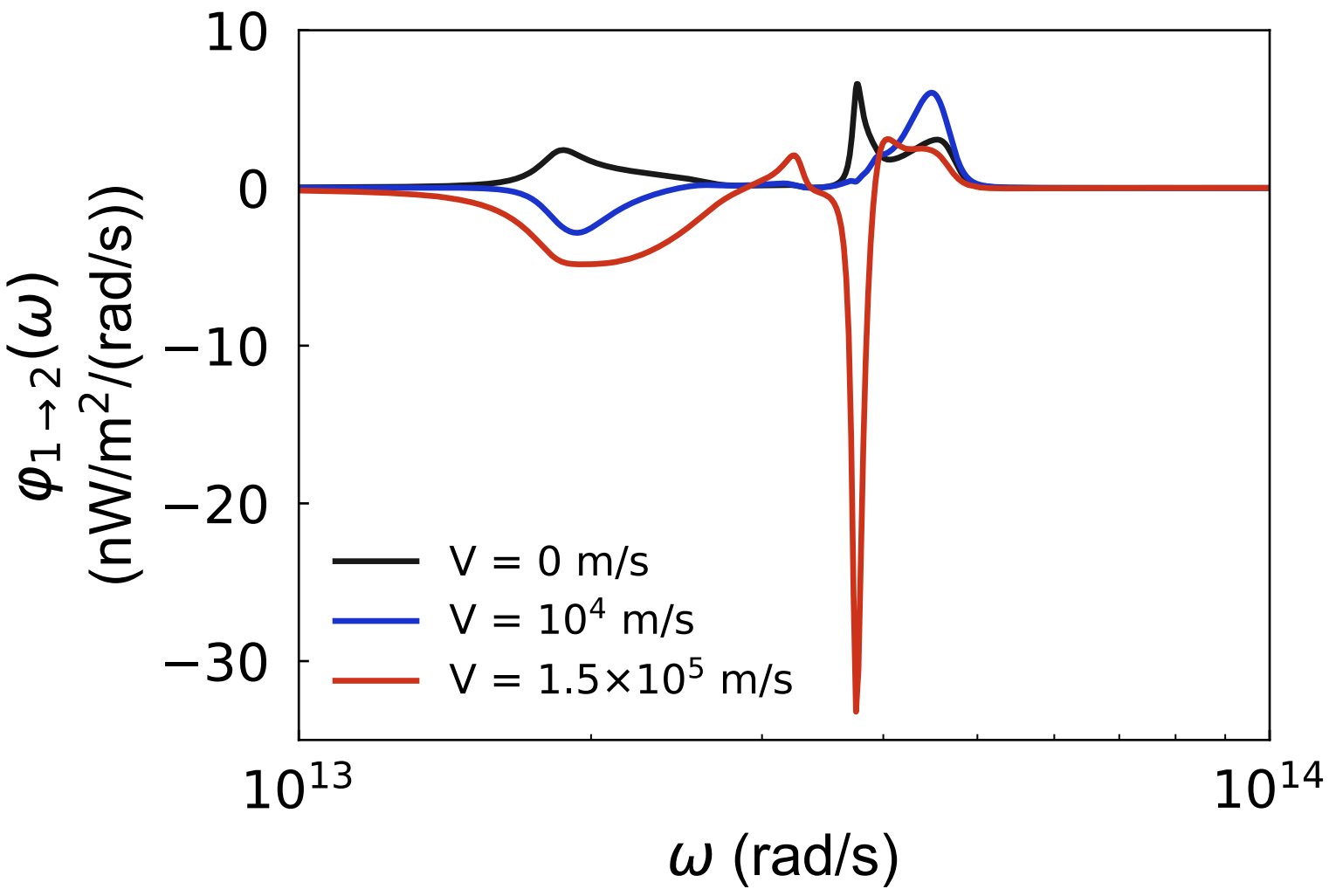

FIG. 5. Spectral radiative heat flux between two slabs under anti-parallel static magnetic fields of 3T for different velocities of relative motion. The other physical parameters are the same as those in Fig. 2.

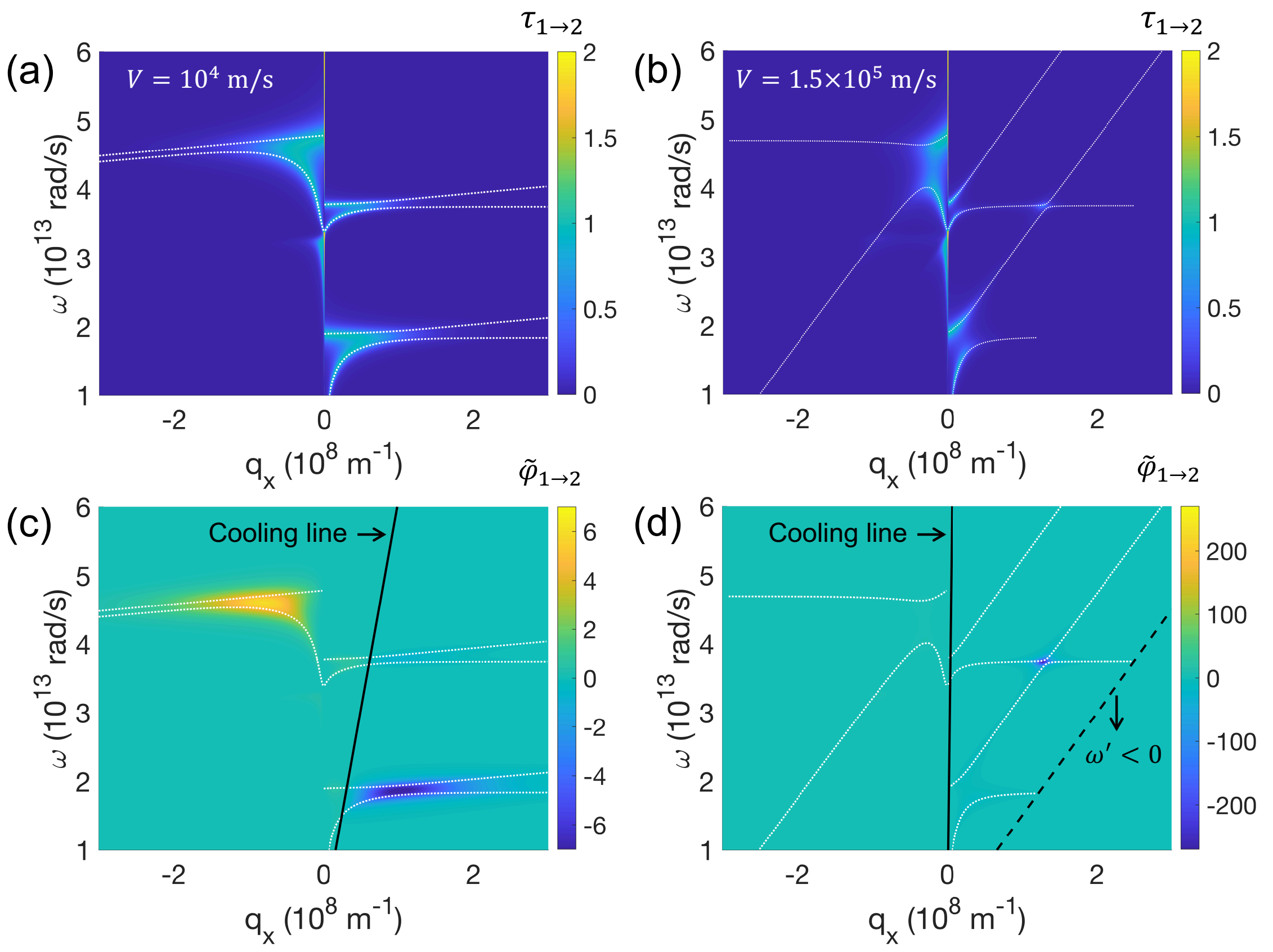

FIG. 6. (a) and (b): Plot of the transmission coefficient $\tau$ as a function of the frequency $\omega$ and $x$-component of the wavevector $q_x$ at $V = 10^4$ m/s and $V = 1.5 \times 10^5$ m/s, respectively. (c) and (d): Plot of mode-resolved radiative heat flux $\tilde{\varphi}_{1\to2}(\omega, q_x, q_y = 0)$ defined in the caption of Fig. 3. The unit is $[10^{-15}\,\frac{\mathrm{W\cdot s}}{\mathrm{rad}}]$. The physical parameters are the same as those in Fig. 2. The black lines in panels (c) and (d) are the constraint for cooling in Eq. (22). The dashed line in panel (d) is the line below which the angular frequency in the co-moving frame is negative, i.e., $\omega' < 0$.

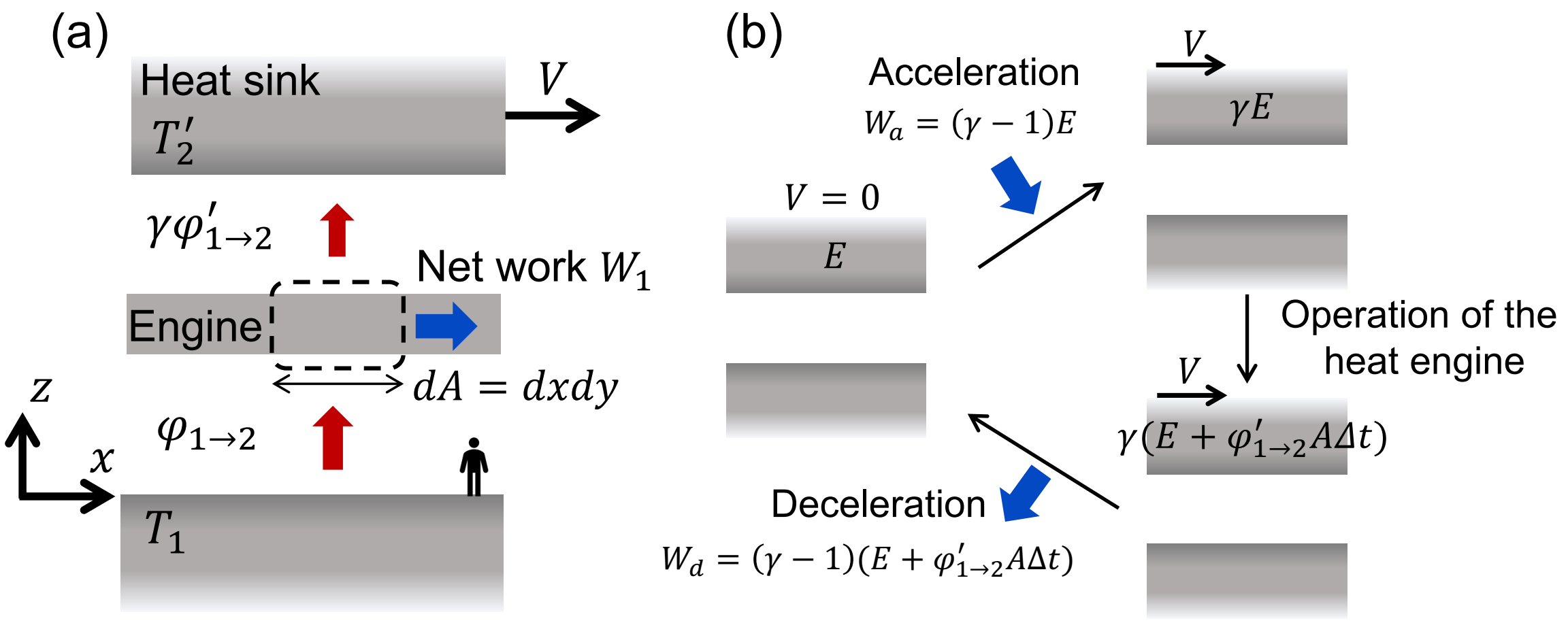

FIG. 7. (a) Application of the first law of relativistic thermodynamics on the control volume enclosed by the dashed line. Slab 2 is hypothetically separated into the engine and heat sink, both of which are moving at the velocity $V$. (b): the evolution of the energy of slab 2 and the work done to slab 2 for acceleration $W_a$ and the work extracted by decelerating slab 2 $W_a$. For both panels, we consider the observer is in the rest frame with slab 1.

Supplementary Information for

# Moving Media as Photonic Heat Engine and Pump

Yoichiro Tsurimaki, Renwen Yu, and Shanhui Fan

Department of Electrical Engineering, Ginzton Laboratory, Stanford University, Stanford, 94305, CA

## 1. Derivation of Radiative Heat Flux and Non-equilibrium Lateral Casimir Force

### 1.1. Total electric and magnetic fields between two slabs with relative motion

We consider that two semi-infinite parallel slabs are separated by a vacuum of gap $d$ and that those two slabs are moving at a velocity $V$ relative to each other as shown in Fig. S1 below. We introduce two coordinate systems. In the rest frame, slab 1 is at rest while slab 2 is moving at the velocity $V$ along the $x$-axis. In the co-moving frame, slab 2 is at rest while slab 1 is moving at the velocity $-V$ along the $x$-axis. For the rest and co-moving frames, we denote $xyz$ and $x'y'z'$ as axes, respectively, as shown in Fig. S1. The assumption of the velocity pointing to the $x$-direction does not lose generality and can be extended to any velocity direction in the $xy$-plane by properly rotating the media. We do not consider the case where the velocity is in the $z$-direction.

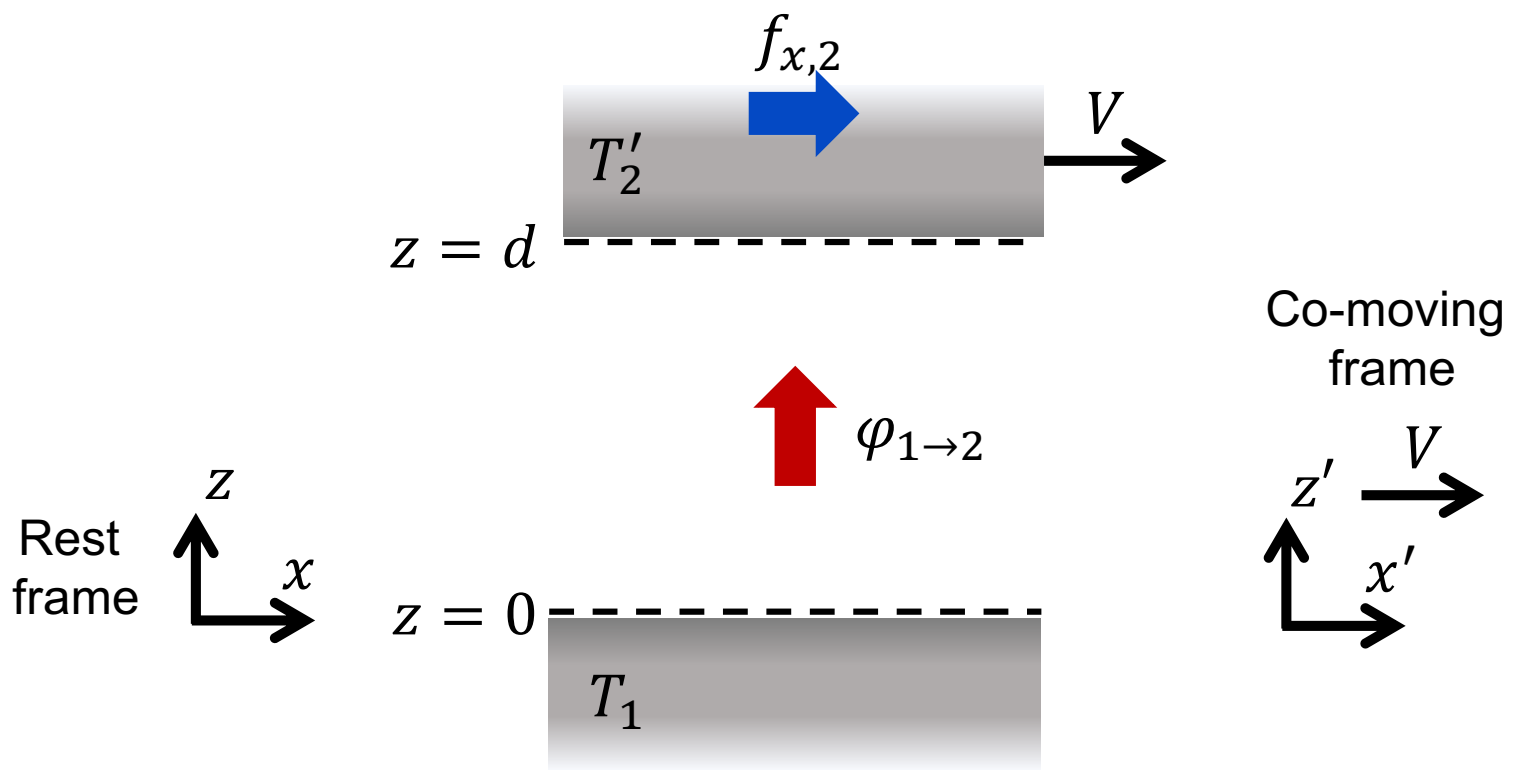

Fig. S1. Schematic of two parallel slabs considered in this work.

E-mail: shanhui@stanford.edu

Since the system is translationally invariant in the $\boldsymbol{R} = (x, y)$ plane, the electromagnetic field can be written down in the Fourier transform as

$$\boldsymbol{E}(\boldsymbol{R}, z, \omega) = \int_{-\infty}^{\infty} dt\, \boldsymbol{E}(\boldsymbol{R}, z, t) e^{i\omega t}, \tag{1.1.1}$$

$$\boldsymbol{E}(\boldsymbol{R}, z, t) = \int_{-\infty}^{\infty} \frac{d\omega}{2\pi} \boldsymbol{E}(\boldsymbol{R}, z, \omega) e^{-i\omega t}, \tag{1.1.2}$$

$$\boldsymbol{E}(\boldsymbol{q}, z, t) = \int_{-\infty}^{\infty} d\boldsymbol{R}\, \boldsymbol{E}(\boldsymbol{R}, z, t) e^{-i\boldsymbol{q}\cdot\boldsymbol{R}}, \tag{1.1.3}$$

$$\boldsymbol{E}(\boldsymbol{R}, z, t) = \int_{-\infty}^{\infty} \frac{d\boldsymbol{q}}{(2\pi)^2} \boldsymbol{E}(\boldsymbol{q}, z, t) e^{i\boldsymbol{q}\cdot\boldsymbol{R}}. \tag{1.1.4}$$

We consider an electric field with an in-plane momentum $\boldsymbol{q} = (q_x, q_y)$ and frequency $\omega$ at the position $z$. It can be written down as a superposition of the forward $(\boldsymbol{q}, k_z)$ and backward $(\boldsymbol{q}, -k_z)$ propagating waves with *s*- and *p*-polarizations. Here $k_z$ is the *z*-component of the wavevector. Apart from the associated coefficients to be determined, the electric field can be written down as:

$$\boldsymbol{E}(\omega, \boldsymbol{q}, z) = \left(v_s \boldsymbol{s} + v_p \boldsymbol{p}_+\right) e^{ik_z z} + \left(w_s \boldsymbol{s} + w_p \boldsymbol{p}_-\right) e^{-ik_z z}, \tag{1.1.5}$$

where $v_j, w_j$ $(j = s, p)$ are the coefficients for the forward and backward waves. $\boldsymbol{s}$ and $\boldsymbol{p}_\pm$ are the unit polarization vectors for the *s*-polarized waves, and forward and backward propagating *p*-polarized waves, respectively. In the rest frame, they are given as:

$$\boldsymbol{s} = \boldsymbol{e_z} \times \frac{\boldsymbol{q}}{q} = \begin{bmatrix} -q_y \\ q_x \\ 0 \end{bmatrix} / q, \quad \boldsymbol{p}_\pm = \boldsymbol{k}_\pm \times \boldsymbol{s} = \begin{bmatrix} \mp q_x k_z \\ \mp q_y k_z \\ q^2 \end{bmatrix} / (k_0 q), \tag{1.1.6}$$

where $k_0 = \frac{\omega}{c}$, $q = |\boldsymbol{q}|$, and $\boldsymbol{k}_\pm = \left[q_x, q_y, \pm k_z\right]$. The magnetic induction field is written down from Faraday's law as

$$\omega\boldsymbol{B}(\omega,\boldsymbol{q},z)=\left(v_s\boldsymbol{k}_+\times\boldsymbol{s}+v_p\boldsymbol{k}_+\times\boldsymbol{p}_+\right)e^{ik_zz}+\left(w_s\boldsymbol{k}_-\times\boldsymbol{s}+w_p\boldsymbol{k}_-\times\boldsymbol{p}_-\right)e^{-ik_zz}. \qquad (1.1.7)$$

We consider the same wave described in the co-moving frame. The Lorentz transformation for the momentum and energy in a vacuum between the rest and co-moving frames is:

$$q'_x=\gamma(q_x-\beta k_0),\qquad q'_y=q_y,\qquad q'_z=q_z,\qquad \omega'=\gamma(\omega-q_xV), \qquad (1.1.8)$$

where $\gamma^{-1}=\sqrt{1-\beta^2}$ and $\beta=\frac{V}{c}$. We use the primes to describe physical quantities in the co-moving frame. The polarization vectors are accordingly transformed as:

$$\boldsymbol{s}'=\boldsymbol{e}'_{\boldsymbol{z}}\times\frac{\boldsymbol{q}'}{q'}=\begin{bmatrix}-q_y\\ q'_x\\ 0\end{bmatrix}/q',\quad \boldsymbol{p}'_\pm=\boldsymbol{k}'_\pm\times\boldsymbol{s}'=\begin{bmatrix}\mp q'_xk_z\\ \mp q_yk_z\\ {q'}^2\end{bmatrix}/(k'_0q'), \qquad (1.1.9)$$

where $k'_0=\frac{\omega'}{c}$ and $q'=|\boldsymbol{q}'|$, and $\boldsymbol{k}'_\pm=\left[q'_x,q_y,\pm k_z\right]$.

The Lorentz transformation of the electromagnetic fields are:

$$\begin{cases}E'_x=E_x\\ E'_y=\gamma\left(E_y-\beta cB_z\right),\\ E'_z=\gamma\left(E_z+\beta cB_y\right)\end{cases} \qquad (1.1.10)$$

$$\begin{cases}B'_x=B_x\\ B'_y=\gamma\left(B_y+\frac{\beta E_z}{c}\right)\\ B'_z=\gamma\left(B_z-\frac{\beta E_y}{c}\right)\end{cases}. \qquad (1.1.11)$$

In the co-moving frame, the electric and magnetic induction fields are similarly written down as:

$$\boldsymbol{E}'(\omega',\boldsymbol{q}',z)=\left(v'_s\boldsymbol{s}'+v'_p\boldsymbol{p}'_+\right)e^{ik_zz}+\left(w'_s\boldsymbol{s}'+w'_p\boldsymbol{p}'_-\right)e^{-ik_zz}, \qquad (1.1.12)$$

$$\omega'\boldsymbol{B}'(\omega',\boldsymbol{q}',z)=\left(v'_s\boldsymbol{k}'_+\times\boldsymbol{s}'+v'_p\boldsymbol{k}'_+\times\boldsymbol{p}'_+\right)e^{ik_zz}+\left(w'_s\boldsymbol{k}'_-\times\boldsymbol{s}'+w'_p\boldsymbol{k}'_-\times\boldsymbol{p}'_-\right)e^{-ik_zz}. \qquad (1.1.13)$$

Note that $z' = z$ and $k_z' = k_z$ in the Lorentz boost along the *x*-direction. In order to find the relations between the coefficients in Eqs. (1.1.5), (1.1.7), (1.1.12), and (1.1.13), we use the fact that the forward propagating waves at $z = 0$ is a superposition of waves that are emitted from slab 1 directly reaching at $z = 0$, and waves that are reflected at the surface of slab 1. This is expressed as

$$\begin{aligned} v_s &= r_1^{ss}(\omega, \boldsymbol{q}) w_s + r_1^{sp}(\omega, \boldsymbol{q}) w_p + E_1^s, \\ v_p &= r_1^{pp}(\omega, \boldsymbol{q}) w_p + r_1^{ps}(\omega, \boldsymbol{q}) w_s + E_1^p, \end{aligned} \tag{1.1.14}$$

where $E_1^{s(p)}$ is the amplitude of the $s(p)$-polarized electric field due to fluctuating currents in slab 1, and $\boldsymbol{E_1} = E_1^s \boldsymbol{s} + E_1^p \boldsymbol{p}_+$. $r_1^{mn}$ $(m, n = s, p)$ is the Fresnel reflection coefficient of slab 1 for *n*-polarized incident field and *m*-polarized reflected field.

Similarly, at $z = z' = d$, in the co-moving frame, we can find the relations as

$$\begin{aligned} w_s' e^{-ik_z d} &= {r'}_2^{ss} v_s' e^{ik_z d} + {r'}_2^{sp} v_p' e^{ik_z d} + {E'}_2^s, \\ w_p' e^{-ik_z d} &= {r'}_2^{pp} v_p' e^{ik_z d} + {r'}_2^{ps} v_s' e^{ik_z d} + {E'}_2^p, \end{aligned} \tag{1.1.15}$$

where we introduced a shorthand notation for the Fresnel reflection coefficient of slab 2 in the co-moving frame, ${r'}_2^{mn} = r_2^{mn}(\omega', \boldsymbol{q}'), (m, n = s, p)$. The condition at $z = d$ is written down in the co-moving frame since the fluctuation-dissipation theorem about the electric field due to fluctuating currents in slab 2, $\boldsymbol{E_2'} = {E'}_2^s \boldsymbol{s'} + {E'}_2^p \boldsymbol{p}'_-$, is written down in the co-moving frame where the local thermodynamic condition is established.

We explicitly write down the electric and magnetic induction fields Eqs. (1.1.5), (1.1.7), (1.1.12), and (1.1.13) as:

$$\begin{aligned} \boldsymbol{E}(\omega, \boldsymbol{q}, z) = &\left\{ v_s \begin{bmatrix} -q_y/q \\ q_x/q \\ 0 \end{bmatrix} + v_p \begin{bmatrix} -q_x k_z \\ -q_y k_z \\ q^2 \end{bmatrix} /(k_0 q) \right\} e^{ik_z z} \\ &+ \left\{ w_s \begin{bmatrix} -q_y/q \\ q_x/q \\ 0 \end{bmatrix} + w_p \begin{bmatrix} q_x k_z \\ q_y k_z \\ q^2 \end{bmatrix} /(k_0 q) \right\} e^{-ik_z z}, \end{aligned} \tag{1.1.16}$$

$$\boldsymbol{E}'(\omega',\boldsymbol{q}',z)=\left\{v_s'\begin{bmatrix}-q_y/q'\\ q_x'/q'\\ 0\end{bmatrix}+v_p'\begin{bmatrix}-q_x'k_z\\ -q_yk_z\\ q'^2\end{bmatrix}/(k_0'q')\right\}e^{ik_zz}$$

$$+\left\{w_s'\begin{bmatrix}-q_y/q'\\ q_x'/q'\\ 0\end{bmatrix}+w_p'\begin{bmatrix}q_x'k_z\\ q_yk_z\\ q'^2\end{bmatrix}/(k_0'q')\right\}e^{-ik_zz}, \qquad (1.1.17)$$

$$\boldsymbol{B}(\omega,\boldsymbol{q},z)=\frac{1}{c}\left\{v_s\begin{bmatrix}-q_xk_z\\ -q_yk_z\\ q^2\end{bmatrix}/(k_0q)+v_p\begin{bmatrix}q_y\\ -q_x\\ 0\end{bmatrix}/q\right\}e^{ik_zz}$$

$$+\frac{1}{c}\left\{w_s\begin{bmatrix}q_xk_z\\ q_yk_z\\ q^2\end{bmatrix}/(k_0q)+w_p\begin{bmatrix}q_y\\ -q_x\\ 0\end{bmatrix}/q\right\}e^{-ik_zz}, \qquad (1.1.18)$$

$$\boldsymbol{B}'(\omega',\boldsymbol{q}',z)=\frac{1}{c}\left\{v_s'\begin{bmatrix}-q_x'k_z\\ -q_yk_z\\ q'^2\end{bmatrix}/(k_0'q')+v_p'\begin{bmatrix}q_y\\ -q_x'\\ 0\end{bmatrix}/q'\right\}e^{ik_zz}$$

$$+\frac{1}{c}\left\{w_s'\begin{bmatrix}q_x'k_z\\ q_yk_z\\ q'^2\end{bmatrix}/(k_0'q')+w_p'\begin{bmatrix}q_y\\ -q_x'\\ 0\end{bmatrix}/q'\right\}e^{-ik_zz}. \qquad (1.1.19)$$

Noting that $e^{ik_zz}$ and $e^{-ik_zz}$ are linearly independent solutions, and the Lorentz transformation $E_x'=E_x$ gives

$$\begin{cases}v_s'\dfrac{q_y}{q'}+v_p'\dfrac{q_x'k_z}{k_0'q'}=v_s\dfrac{q_y}{q}+v_p\dfrac{q_xk_z}{k_0q}\\ w_s'\dfrac{q_y}{q'}+w_p'\dfrac{q_x'k_z}{k_0'q'}=w_s\dfrac{q_y}{q}+w_p\dfrac{q_xk_z}{k_0q}\end{cases}. \qquad (1.1.20)$$

Another condition $E_z'=\gamma\left(E_z+\beta cB_y\right)$ gives

$$\begin{cases}v_p'\dfrac{q'}{k_0'}=\gamma\left(v_p\dfrac{q}{k_0}-v_s\beta\dfrac{q_yk_z}{k_0q}-v_p\beta\dfrac{q_x}{q}\right)\\ w_p'\dfrac{q'}{k_0'}=\gamma\left(w_p\dfrac{q}{k_0}+w_s\beta\dfrac{q_yk_z}{k_0q}-w_p\beta\dfrac{q_x}{q}\right)\end{cases}. \qquad (1.1.21)$$

Solving Eqs. (1.1.20) and (1.1.21) for $v_p'$ and $w_p'$, we have:

$$v_p' = \frac{k_0'\gamma}{k_0 q q'}\{-\beta q_y k_z v_s + (q^2 - \beta q_x k_0) v_p\}, \tag{1.1.22}$$

$$w_p' = \frac{k_0'\gamma}{k_0 q q'}\{\beta q_y k_z w_s + (q^2 - \beta q_x k_0) w_p\}. \tag{1.1.23}$$

Similarly, we use the Lorentz transformation for the magnetic induction fields Eqs. (1.1.11), (1.1.18), and (1.1.19), we have the relation between the *s*-polarized coefficients

$$v_s' = \frac{k_0'\gamma}{k_0 q q'}\{\beta q_y k_z v_p + (q^2 - \beta q_x k_0) v_s\}, \tag{1.1.24}$$

$$w_s' = \frac{k_0'\gamma}{k_0 q q'}\{-\beta q_y k_z w_p + (q^2 - \beta q_x k_0) w_s\}. \tag{1.1.25}$$

We can compactly write down Eqs. (1.1.22)-(1.1.25) as:

$$\begin{bmatrix} v_s' \\ v_p' \end{bmatrix} = L \begin{bmatrix} v_s \\ v_p \end{bmatrix}, \qquad \begin{bmatrix} w_s' \\ w_p' \end{bmatrix} = L^T \begin{bmatrix} w_s \\ w_p \end{bmatrix}, \tag{1.1.26}$$

where

$$L = \frac{k_0'\gamma}{k_0 q q'} \begin{bmatrix} q^2 - \beta k_0 q_x & \beta k_z q_y \\ -\beta k_z q_y & q^2 - \beta k_0 q_x \end{bmatrix}. \tag{1.1.27}$$

The matrix $L$ transforms the coefficients for the forward propagating waves from the rest frame to the co-moving frame via the Lorentz transformation and its transpose transforms the coefficients of the backward propagating waves from the rest to the co-moving frames.

Noting that we can write

$$q q' = \gamma \sqrt{(q^2 - \beta k_0 q_x)^2 + \left(\beta q_y k_z\right)^2}, \tag{1.1.28}$$

we can find

$$L^{-1} = \left(\frac{k_0}{k_0'}\right)^2 L^T, \tag{1.1.29}$$

because $\left(\frac{k_0'}{k_0}\right)^2 = \det L$. We note that $\det L = 0$ when the wave at the frequency $\omega > 0$ in the rest frame is Lorentz boosted to the velocity $V = \frac{\omega}{q_x}$ at which the frequency in the co-moving frame is $\omega' = 0$. However, such static fields in the co-moving frame do not contribute to energy and momentum transfer. Thus, the inverse of the matrix $L$ exists and $\tilde{R}_2$ is well defined as long as we do not consider the Lorentz boost to the frame where $\omega' = 0$. However, our formulas Eqs. (1) and (2) in the main manuscript produce null heat transfer and shear stress for modes with $\omega' = 0$, and the exclusion of such Lorentz boost does not affect our results.

Solving Eqs. (1.1.14), (1.1.15), and (1.1.26), we can write down the coefficients associated with the electric fields in the rest frame in terms of the amplitudes of the fluctuating electric fields due to slabs 1 and 2. The results are

$$\begin{bmatrix} v_s \\ v_p \end{bmatrix} = \frac{1}{D_{12}}\left(\frac{k_0'}{k_0}\right)^2 \begin{bmatrix} E_1^s \\ E_1^p \end{bmatrix} + \frac{1}{D_{12}} R_1 L e^{ik_z d} \begin{bmatrix} E'^s_2 \\ E'^p_2 \end{bmatrix}, \tag{1.1.30}$$

$$\begin{bmatrix} w_s \\ w_p \end{bmatrix} = L \frac{1}{D_{21}} R_2' L e^{i2k_z d} \begin{bmatrix} E_1^s \\ E_1^p \end{bmatrix} + L \frac{1}{D_{21}} e^{ik_z d} \begin{bmatrix} E'^s_2 \\ E'^p_2 \end{bmatrix}, \tag{1.1.31}$$

where $R_1$and $R_2'$ are the reflection matrices whose elements are the Fresnel reflection coefficients. They are defined in Eq. (3) in the main text. $D_{12}$ and $D_{21}$ describe the multi-reflection of waves between the two slabs and are given as

$$D_{12} = \left(\frac{k_0'}{k_0}\right)^2 I - R_1 L R_2' L e^{i2k_z d}, \tag{1.1.32}$$

$$D_{21} = \left(\frac{k_0'}{k_0}\right)^2 I - R_2' L R_1 L e^{i2k_z d}, \tag{1.1.33}$$

where $I$ is the identity matrix. With Eqs. (1.1.30) and (1.1.31) in hand, we are ready to compute the radiative heat flux and shear stress between the two slabs.

### 1.2. Radiative heat flux between two slabs

The radiative heat flux between the two slabs is expressed by the ensemble-averaged Poynting vector

$$\boldsymbol{q}(\boldsymbol{r},t) = \langle \boldsymbol{E}(\boldsymbol{r},t) \times \boldsymbol{H}(\boldsymbol{r},t)\rangle_E, \tag{1.2.1}$$

where $\langle \quad \rangle_E$ denotes the ensemble average. In the spatial and temporal Fourier domain, we have

$$\boldsymbol{q}(\boldsymbol{r},t) = \int_{-\infty}^{\infty} \frac{d\omega}{2\pi} \int \frac{d\boldsymbol{q}}{(2\pi)^2} \int_{-\infty}^{\infty} \frac{d\omega''}{2\pi} \int \frac{d\boldsymbol{q}''}{(2\pi)^2} \times \\ \langle \boldsymbol{E}(\boldsymbol{q},z,\omega) \times \boldsymbol{H}(\boldsymbol{q}'',z,\omega'')\rangle_E e^{-i(\omega+\omega'')t} e^{i(\boldsymbol{q}+\boldsymbol{q}'')\cdot\boldsymbol{R}}. \tag{1.2.2}$$

The stationarity, as well as the translational invariance of the system, results in

$$\langle \boldsymbol{E}(\boldsymbol{q},z,\omega) \times \boldsymbol{H}(\boldsymbol{q}'',z,\omega'')\rangle = \langle \boldsymbol{E}(\boldsymbol{q},z,\omega) \times \boldsymbol{H}(\boldsymbol{q}'',z,\omega'')\rangle \delta(\omega+\omega'')\delta(\boldsymbol{q}+\boldsymbol{q}''). \tag{1.2.3}$$

Note that $\boldsymbol{q}''$ and $\omega''$ are the quantities in the rest frame. In fact, these delta functions naturally appear from the fluctuation-dissipation theorem. Thus, the heat flux is no longer a function of $\boldsymbol{R}$ and time and we have

$$\boldsymbol{\varphi} = \int_{-\infty}^{\infty} \frac{d\omega}{(2\pi)^2} \int \frac{d\boldsymbol{q}}{(2\pi)^4} \langle \boldsymbol{E}(\boldsymbol{q},z,\omega) \times \boldsymbol{H}^*(\boldsymbol{q},z,\omega)\rangle_E \\ = \int_0^{\infty} \frac{d\omega}{2\pi} \int \frac{d\boldsymbol{q}}{(2\pi)^4} \frac{1}{2\pi} \langle \boldsymbol{E}(\boldsymbol{q},z,\omega) \times \boldsymbol{H}^*(\boldsymbol{q},z,\omega) + \boldsymbol{E}^*(\boldsymbol{q},z,\omega) \times \boldsymbol{H}(\boldsymbol{q},z,\omega)\rangle_E, \tag{1.2.4}$$

Since we are interested in the $z$-component of the Poynting vector, we have

$$\varphi_z = \int_0^{\infty} \frac{d\omega}{2\pi} \int \frac{d\boldsymbol{q}}{(2\pi)^4} \frac{1}{\pi} \langle \boldsymbol{E}(\boldsymbol{q}, z, \omega) \times \boldsymbol{H}^*(\boldsymbol{q}, z, \omega) \rangle \cdot \boldsymbol{e}_z|_{z=0}, \tag{1.2.5}$$

Inserting Eqs. (1.1.5) and (1.1.7), and using $\boldsymbol{B} = \mu_0 \boldsymbol{H}$, we find

$$\begin{aligned} &\langle \boldsymbol{E}(\boldsymbol{q}, z, \omega) \times \boldsymbol{H}^*(\boldsymbol{q}, z, \omega) + \boldsymbol{E}^*(\boldsymbol{q}, z, \omega) \times \boldsymbol{H}(\boldsymbol{q}, z, \omega) \rangle_E \cdot \boldsymbol{e}_z|_{z=0} \\ &= \frac{2}{c\mu_0} \mathrm{Tr} \begin{bmatrix} \frac{\mathrm{Re}[k_z]}{k_0} \left\{ \langle \begin{bmatrix} v_s \\ v_p \end{bmatrix} [v_s^* \quad v_p^*] \rangle_S - \langle \begin{bmatrix} w_s \\ w_p \end{bmatrix} [w_s^* \quad w_p^*] \rangle_S \right\} \\ +i \frac{\mathrm{Im}[k_z]}{k_0} \left\{ \langle \begin{bmatrix} v_s \\ v_p \end{bmatrix} [w_s^* \quad w_p^*] \rangle_S - \langle \begin{bmatrix} w_s \\ w_p \end{bmatrix} [v_s^* \quad v_p^*] \rangle_S \right\} \end{bmatrix}, \end{aligned} \tag{1.2.6}$$

where the symmetrized correlation function is defined as $\langle AB \rangle_S = \frac{1}{2} \langle AB + BA \rangle_E$. By inserting Eqs. (1.1.30) and (1.1.31), this can be further proceeded as

$$\begin{aligned} &\frac{2}{c\mu_0} \frac{\mathrm{Re}[k_z]}{k_0} \mathrm{Tr} \begin{bmatrix} \left( \frac{k_0'}{k_0} \right)^2 \left\{ \left( \frac{k_0'}{k_0} \right)^2 I - L^\dagger R_2'^\dagger R_2' L \right\} \frac{1}{D_{12}} \langle \boldsymbol{E}_1 \boldsymbol{E}_1^\dagger \rangle_S \frac{1}{D_{12}^\dagger} \\ - \left\{ \left( \frac{k_0'}{k_0} \right)^2 I - L^\dagger R_1^\dagger R_1 L \right\} \frac{1}{D_{21}} \langle \boldsymbol{E}_2' \boldsymbol{E}_2'^\dagger \rangle_S \frac{1}{D_{21}^\dagger} \end{bmatrix} \\ &+ \frac{2i}{c\mu_0} \frac{\mathrm{Im}[k_z]}{k_0} \mathrm{Tr} \begin{bmatrix} \left( \frac{k_0'}{k_0} \right)^2 L \left( R_2'^\dagger - R_2' \right) L \frac{1}{D_{12}} \langle \boldsymbol{E}_1 \boldsymbol{E}_1^\dagger \rangle_S \frac{1}{D_{12}^\dagger} \\ -L \left( R_1 - R_1^\dagger \right) L \frac{1}{D_{21}} \langle \boldsymbol{E}_2' \boldsymbol{E}_2'^\dagger \rangle_S \frac{1}{D_{21}^\dagger} \end{bmatrix} e^{-2\mathrm{Im}[k_z]d}, \end{aligned} \tag{1.2.7}$$

To derive Eq. (1.2.7), we used the identities $\frac{1}{D_{21}} R_2' L = R_2' L \frac{1}{D_{12}}$ and $\frac{1}{D_{12}} R_1 L = R_1 L \frac{1}{D_{21}}$ as well as the fact that $L$ is a real matrix for $k_z \in \mathbb{R}$ and is Hermitian when $k_z$ is purely imaginary.

From here, we separately consider the propagative waves where $q < \omega/c$ and evanescent waves $q > \omega/c$. First, we consider the propagative waves and the terms associated with $\langle \boldsymbol{E}_1 \boldsymbol{E}_1^\dagger \rangle_S$. Since $k_z$ is purely real, only the first term in Eq. (1.2.7) is non-zero. The electric field correlations are given by the fluctuation-dissipation theorem [1] as

$$\langle \boldsymbol{E}_1 \boldsymbol{E}_1^\dagger \rangle_S = \pi(2\pi)^2 \sqrt{\frac{\mu_0}{\varepsilon_0}} \frac{k_0}{k_z} \Theta(\omega, T_1)\left(I - R_1 R_1^\dagger\right), \tag{1.2.8}$$

$$\langle \boldsymbol{E}'_2 \boldsymbol{E}'^\dagger_2 \rangle_S = \pi(2\pi)^2 \sqrt{\frac{\mu_0}{\varepsilon_0}} \frac{k'_0}{k_z} \Theta(\omega', T'_2)\left(I - R'_2 R'^\dagger_2\right), \tag{1.2.9}$$

where $\Theta(\omega, T) = \frac{\hbar\omega}{2} + \frac{\hbar\omega}{e^{\frac{\hbar\omega}{k_B T}} - 1}$. The difference about the factor $\pi$ in this work and that in [1] originates from the different definitions of the Fourier transform. Inserting these two in the first term in Eq. (1.2.7), we obtain the propagative component of the net radiative heat flux from slabs 1 to 2 as

$$\varphi_{1\to 2}^{prop} = \int_0^\infty \frac{d\omega}{2\pi} \int_{q<\frac{\omega}{c}} \frac{d\boldsymbol{q}}{(2\pi)^2} \left[ \frac{\hbar\omega}{e^{\frac{\hbar\omega}{k_B T_1}} - 1} \tau_{1\to 2}^{prop}(\omega, \boldsymbol{q}) - \frac{\hbar\omega}{e^{\frac{\hbar\gamma(\omega - q_x V)}{k_B T'_2}} - 1} \tau_{2\to 1}^{prop}(\omega, \boldsymbol{q}) \right], \tag{1.2.10}$$

where

$$\tau_{1\to 2}^{prop}(\omega, \boldsymbol{q}) = \mathrm{Tr} \left(\frac{k'_0}{k_0}\right)^2 \left[ \left( \left(\frac{k'_0}{k_0}\right)^2 I - L^\dagger R'^\dagger_2 R'_2 L \right) \frac{1}{D_{12}} \left(I - R_1 R_1^\dagger\right) \frac{1}{D_{12}^\dagger} \right], \tag{1.2.11}$$

and

$$\tau_{2\to 1}^{prop}(\omega, \boldsymbol{q}) = \mathrm{Tr} \left(\frac{k'_0}{k_0}\right)^2 \left[ \left( \left(\frac{k'_0}{k_0}\right)^2 I - L^\dagger R_1^\dagger R_1 L \right) \frac{1}{D_{21}} \left(I - R'_2 R'^\dagger_2\right) \frac{1}{D_{21}^\dagger} \right]. \tag{1.2.12}$$

The zero-point energy terms in Eq. (1.2.10) are canceled out due to the identity $\tau_{1\to 2} = \tau_{2\to 1}$ that we prove in the later section.

We next consider the contribution from the evanescent waves where $k_z = i\kappa_z$ ($\kappa_z \in \mathbb{R}$) is purely imaginary. The electric field correlations are given by the fluctuation-dissipation theorem as

$$\langle \boldsymbol{E}_1 \boldsymbol{E}_1^\dagger \rangle_S = \pi(2\pi)^2 \sqrt{\frac{\mu_0}{\varepsilon_0}} \frac{k_0}{i\kappa_z} \Theta(\omega, T_1)\left(R_1 - R_1^\dagger\right). \tag{1.2.13}$$

$$\langle \boldsymbol{E}'_2 \boldsymbol{E}'^{\dagger}_2 \rangle_S = \pi (2\pi)^2 \sqrt{\frac{\mu_0}{\varepsilon_0}} \frac{k'_0}{i\kappa_z} \Theta(\omega', T'_2) \left( R'_2 - R'^{\dagger}_2 \right). \tag{1.2.14}$$

Therefore, the evanescent component of the net radiative heat flux from slabs 1 to 2 is expressed as

$$\varphi_{1\to 2}^{evan} = \int_0^\infty \frac{d\omega}{2\pi} \int_{q>\frac{\omega}{c}} \frac{d\boldsymbol{q}}{(2\pi)^2} \left[ \frac{\hbar\omega}{e^{\frac{\hbar\omega}{k_B T_1}} - 1} \tau_{1\to 2}^{evan}(\omega, \boldsymbol{q}) - \frac{\hbar\omega}{e^{\frac{\hbar\gamma(\omega - q_x V)}{k_B T'_2}} - 1} \tau_{2\to 1}^{evan}(\omega, \boldsymbol{q}) \right], \tag{1.2.15}$$

where

$$\tau_{1\to 2}^{evan}(\omega, \boldsymbol{q}) = \mathrm{Tr} \left( \frac{k'_0}{k_0} \right)^2 \left[ L(R'^{\dagger}_2 - R'_2) L \frac{1}{D_{12}} \left( R_1 - R_1^{\dagger} \right) \frac{1}{D_{12}^{\dagger}} \right] e^{-2\kappa_z d}, \tag{1.2.16}$$

and

$$\tau_{2\to 1}^{evan}(\omega, \boldsymbol{q}) = \mathrm{Tr} \left( \frac{k'_0}{k_0} \right)^2 \left[ L(R_1^{\dagger} - R_1) L \frac{1}{D_{21}} \left( R'_2 - R'^{\dagger}_2 \right) \frac{1}{D_{21}^{\dagger}} \right] e^{-2\kappa_z d}. \tag{1.2.17}$$

In Eqs. (1.2.11) and (1.2.16), the matrix sandwiched by $L$ such as $LR'_2L$ appear. Using Eq. (1.1.29),

$$\left( \frac{k_0}{k'_0} \right)^2 LR'_2L = LR'_2LL^{-1}(L^T)^{-1} = LR'_2(L^T)^{-1} \equiv \tilde{R}_2. \tag{1.2.18}$$

Here, we defined a matrix

$$\tilde{R}_2 \equiv LR'_2(L^T)^{-1} = (L^T)^{-1}R'_2L. \tag{1.2.19}$$

We can interpret $\tilde{R}_2$ as the reflection matrix of a moving slab measured in the rest frame. $\tilde{R}_2$ is given as the product of the Lorentz transformation for the forward propagating waves from the rest frame to the co-moving frame ($L$), the reflection matrix of slab 2 in the co-moving frame ($R'_2$) and the Lorentz transformation for the backward propagating waves from the co-moving frame to the rest frame ($(L^T)^{-1}$). By using the identities Eq. (1.2.18) and $\tau_{1\to2}^{prop(evan)} = \tau_{2\to1}^{prop(evan)}$ which we prove later, we can write down the radiative heat flux as well as the transmission coefficient as shown in Eqs. (1) and (2) in the main text.

### 1.3. Shear stress between two slabs

The shear stress and radiation pressure between the two slabs can be determined from the Maxwell's stress tensor

$$\tau_{ij}(\boldsymbol{r},t) = \varepsilon_0\left[E_iE_j + c^2B_iB_j + \delta_{ij}(|\boldsymbol{E}|^2 + c^2|\boldsymbol{B}|^2)\right], \tag{1.3.1}$$

where $\delta_{ij}$ is the Kronecker's delta and $i,j = x,y,z$. The shear stress and radiation pressure are

$$\boldsymbol{f} = \frac{1}{A}\int dA\boldsymbol{n}\cdot\tau = \frac{1}{A}\int dA\begin{bmatrix} n_i\tau_{ix} \\ n_i\tau_{iy} \\ n_i\tau_{iz}\end{bmatrix}, \tag{1.3.2}$$

where $\boldsymbol{n}$ is the normal vector to the surface $A$ of the slab on which the force is calculated. We consider the shear stress on slab 1 in the $x$-direction. It is given as

$$f_{x,1} = \frac{1}{A}\int dA\,\langle\tau_{zx}\rangle = \frac{1}{A}\varepsilon_0\int dA\,[\langle E_zE_x\rangle_E + c^2\langle B_zB_x\rangle_E]|_{z=0}. \tag{1.3.3}$$

The continuous translational invariance of the system makes the integrand in Eq. (1.3.3) position independent, and we can ignore the integral over the area. In the Fourier domain, we have

$$f_{x,1} = \varepsilon_0\int_0^\infty\frac{d\omega}{2\pi}\int\frac{d\boldsymbol{q}}{(2\pi)^4}\frac{1}{2\pi}\times \left[\begin{matrix}\langle E_z(\omega,\boldsymbol{q},z=0)E_x^*(\omega,\boldsymbol{q},z=0)\rangle_E + \langle E_z^*(\omega,\boldsymbol{q},z=0)E_x(\omega,\boldsymbol{q},z=0)\rangle_E \\ +c^2\langle B_z(\omega,\boldsymbol{q},z=0)B_x^*(\omega,\boldsymbol{q},z=0)\rangle_E + c^2\langle B_z^*(\omega,\boldsymbol{q},z=0)B_x(\omega,\boldsymbol{q},z=0)\rangle_E\end{matrix}\right]. \tag{1.3.4}$$

Inserting Eqs. (1.1.5) and (1.1.7), we obtain

$$f_{x,1} = \varepsilon_0 \int_0^\infty \frac{d\omega}{2\pi} \int \frac{d\boldsymbol{q}}{(2\pi)^4} \frac{1}{\pi} \times$$
$$\left\{ \begin{array}{l} \dfrac{q_x \mathrm{Re}[k_z]}{k_0^2} \mathrm{Tr}\left[ -\langle \begin{bmatrix} v_s \\ v_p \end{bmatrix} [v_s^* \quad v_p^*] \rangle_S + \langle \begin{bmatrix} w_s \\ w_p \end{bmatrix} [w_s^* \quad w_p^*] \rangle_S \right] \\ + \dfrac{q_x i \mathrm{Im}[k_z]}{k_0^2} \mathrm{Tr}\left[ -\langle \begin{bmatrix} v_s \\ v_p \end{bmatrix} [w_s^* \quad w_p^*] \rangle_S + \langle \begin{bmatrix} w_s \\ w_p \end{bmatrix} [v_s^* \quad v_p^*] \rangle_S \right] \end{array} \right\}. \tag{1.3.5}$$

Following similar calculations of the radiative heat flux, we obtain the shear stress on the slab 1 as

$$f_{x,1}^{prop} = \int_0^\infty \frac{d\omega}{2\pi} \int_{q<\frac{\omega}{c}} \frac{d\boldsymbol{q}}{(2\pi)^2} \left[ \frac{-\hbar q_x}{e^{\frac{\hbar\omega}{k_B T_1}} - 1} \tau_{1\to2}^{prop}(\omega, \boldsymbol{q}) - \frac{-\hbar q_x}{e^{\frac{\hbar\gamma(\omega - q_x V)}{k_B T_2'}} - 1} \tau_{2\to1}^{prop}(\omega, \boldsymbol{q}) \right], \tag{1.3.6}$$

$$f_{x,1}^{evan} = \int_0^\infty \frac{d\omega}{2\pi} \int_{q>\frac{\omega}{c}} \frac{d\boldsymbol{q}}{(2\pi)^2} \left[ \frac{-\hbar q_x}{e^{\frac{\hbar\omega}{k_B T_1}} - 1} \tau_{1\to2}^{evan}(\omega, \boldsymbol{q}) - \frac{-\hbar q_x}{e^{\frac{\hbar\gamma(\omega - q_x V)}{k_B T_2'}} - 1} \tau_{2\to1}^{evan}(\omega, \boldsymbol{q}) \right]. \tag{1.3.7}$$

The shear stress on slab 2 can be calculated by evaluating the electromagnetic waves at $z = d$. Since the normal vector points in the opposite direction, the force direction becomes opposite

$$f_{x,2} = -f_{x,1}. \tag{1.3.8}$$

## 2. Proof of $\tau_{1\to2}(\omega, \boldsymbol{q}; V) = \tau_{2\to1}(\omega, \boldsymbol{q}; V)$

In this section, we prove the equality $\tau_{1\to2} = \tau_{2\to1}$ for both propagative and evanescent waves. We begin with the case of propagative waves. We expand Eq. (1.2.11)

$$
\begin{aligned}
\tau_{1\to 2}^{prop} &= \mathrm{Tr}\left(\frac{k_0'}{k_0}\right)^2 \left[\left(\left(\frac{k_0'}{k_0}\right)^2 I - L^\dagger R_2'^\dagger R_2' L\right)\frac{1}{D_{12}}\left(I - R_1 R_1^\dagger\right)\frac{1}{D_{12}^\dagger}\right] \\
&= \mathrm{Tr}\left(\frac{k_0'}{k_0}\right)^2 \begin{bmatrix} \left(\frac{k_0'}{k_0}\right)^2 \frac{1}{D_{12}}\frac{1}{D_{12}^\dagger} - \left(\frac{k_0'}{k_0}\right)^2 \frac{1}{D_{12}} R_1 R_1^\dagger \frac{1}{D_{12}^\dagger} \\ -L^\dagger R_2'^\dagger R_2' L \frac{1}{D_{12}}\frac{1}{D_{12}^\dagger} + L^\dagger R_2'^\dagger R_2' L \frac{1}{D_{12}} R_1 R_1^\dagger \frac{1}{D_{12}^\dagger} \end{bmatrix}.
\end{aligned} \tag{2.1}
$$

We try to rewrite the four terms Eq. (2.1). To begin with, the second term in Eq. (2.1) can be written as

$$
\mathrm{Tr}\left(\frac{k_0'}{k_0}\right)^4 \left[\frac{1}{D_{12}} R_1 R_1^\dagger \frac{1}{D_{12}^\dagger}\right] = \mathrm{Tr}\left(\frac{k_0'}{k_0}\right)^2 \left[R_1 L \frac{1}{D_{21}}\frac{1}{D_{21}^\dagger} L^\dagger R_1^\dagger\right], \tag{2.2}
$$

where we used Eq. (1.1.29), the reality of the $L$ matrix for the propagative waves, the identity $R_1 L \frac{1}{D_{21}} = \frac{1}{D_{12}} R_1 L$. Similarly, the third term in Eq. (2.1) can be rewritten as

$$
\mathrm{Tr}\left(\frac{k_0'}{k_0}\right)^2 \left[L^\dagger R_2'^\dagger R_2' L \frac{1}{D_{12}}\frac{1}{D_{12}^\dagger}\right] = \mathrm{Tr}\left(\frac{k_0'}{k_0}\right)^4 \left[\frac{1}{D_{21}} R_2' R_2'^\dagger \frac{1}{D_{21}^\dagger}\right]. \tag{2.3}
$$

In the following, we use the identities

$$
\begin{aligned}
\frac{1}{D_{12}} &= \left(\frac{k_0}{k_0'}\right)^2 \left[I + R_1 L \frac{1}{D_{21}} R_2' L e^{i2k_z d}\right], \\
\frac{1}{D_{21}} &= \left(\frac{k_0}{k_0'}\right)^2 \left[I + R_2' L \frac{1}{D_{12}} R_1 L e^{i2k_z d}\right].
\end{aligned} \tag{2.4}
$$

The first term in Eq. (2.1) is

$$\mathrm{Tr}\left(\frac{k_0'}{k_0}\right)^4\left[\frac{1}{D_{12}}\frac{1}{D_{12}^\dagger}\right]=\mathrm{Tr}\left[\left[I+R_1L\frac{1}{D_{21}}R_2'Le^{i2k_zd}\right]\left[I+L^\dagger R'^\dagger_2\frac{1}{D_{21}^\dagger}L^\dagger R_1^\dagger e^{-i2k_zd}\right]\right]$$

$$=\mathrm{Tr}\left[I+\frac{1}{D_{12}^\dagger}L^\dagger R'^\dagger_2L^\dagger R_1^\dagger e^{-i2k_zd}+\frac{1}{D_{12}}R_1LR_2'Le^{i2k_zd}+\left(\frac{k_0'}{k_0}\right)^2R_1L\frac{1}{D_{21}}R_2'R'^\dagger_2\frac{1}{D_{21}^\dagger}L^\dagger R_1^\dagger\right]. \quad (2.5)$$

If we combine Eqs. (2.5) and the fourth term in (2.1), we obtain

$$\mathrm{Tr}\begin{bmatrix} I+\frac{1}{D_{12}^\dagger}L^\dagger R'^\dagger_2L^\dagger R_1^\dagger e^{-i2k_zd}+\frac{1}{D_{12}}R_1LR_2'Le^{i2k_zd} \\ +\left(\frac{k_0'}{k_0}\right)^2R_1L\frac{1}{D_{21}}R_2'R'^\dagger_2\frac{1}{D_{21}^\dagger}L^\dagger R_1^\dagger+\left(\frac{k_0'}{k_0}\right)^2L^\dagger R'^\dagger_2R_2'L\frac{1}{D_{12}}R_1R_1^\dagger\frac{1}{D_{12}^\dagger}\end{bmatrix}$$

$$=\mathrm{Tr}\left[\left(\frac{k_0'}{k_0}\right)^4\frac{1}{D_{21}}\frac{1}{D_{21}^\dagger}+\left(\frac{k_0'}{k_0}\right)^2R_1L\frac{1}{D_{21}}R_2'R'^\dagger_2\frac{1}{D_{21}^\dagger}L^\dagger R_1^\dagger\right], \quad (2.6)$$

Finally, if we combine Eqs. (2.2), (2.3), and (2.6), we obtain

$$\tau_{1\to2}^{prop}=\mathrm{Tr}\left(\frac{k_0'}{k_0}\right)^2\begin{bmatrix}\left(\frac{k_0'}{k_0}\right)^2\frac{1}{D_{21}}\frac{1}{D_{21}^\dagger}+R_1L\frac{1}{D_{21}}R_2'R'^\dagger_2\frac{1}{D_{21}^\dagger}L^\dagger R_1^\dagger \\ -R_1L\frac{1}{D_{21}}\frac{1}{D_{21}^\dagger}L^\dagger R_1^\dagger-\frac{1}{D_{21}}R_2'R'^\dagger_2\frac{1}{D_{21}^\dagger}\end{bmatrix}$$

$$=\mathrm{Tr}\left(\frac{k_0'}{k_0}\right)^2\left[\left(\left(\frac{k_0'}{k_0}\right)^2I-L^\dagger R_1^\dagger R_1L\right)\frac{1}{D_{21}}\left(I-R_2'R'^\dagger_2\right)\frac{1}{D_{21}^\dagger}\right]=\tau_{2\to1}^{prop}. \quad (2.7)$$

Thus, we showed the identity for the propagative waves.

Next, we show the identity for the evanescent waves. We expand Eq. (1.2.16)

$$\tau_{1\to2}^{evan}=\mathrm{Tr}\left(\frac{k_0'}{k_0}\right)^2\begin{bmatrix}LR_2'^\dagger L\frac{1}{D_{12}}R_1\frac{1}{D_{12}^\dagger}-LR_2'^\dagger L\frac{1}{D_{12}}R_1^\dagger\frac{1}{D_{12}^\dagger} \\ -LR_2'L\frac{1}{D_{12}}R_1\frac{1}{D_{12}^\dagger}+LR_2'L\frac{1}{D_{12}}R_1^\dagger\frac{1}{D_{12}^\dagger}\end{bmatrix}e^{-2\kappa_zd}. \quad (2.8)$$

By using the identity $L = L^\dagger$, which is only valid for purely imaginary $k_z$, the first term in Eq. (2.8) can be rewritten as

$$\mathrm{Tr}\left(\frac{k_0'}{k_0}\right)^2 \left[LR_2'^{\dagger}L\frac{1}{D_{12}}R_1\frac{1}{D_{12}^{\dagger}}\right]e^{-2\kappa_z d} = \mathrm{Tr}\left(\frac{k_0'}{k_0}\right)^2 \left[LR_1L\frac{1}{D_{21}}R_2'^{\dagger}\frac{1}{D_{21}^{\dagger}}\right]e^{-2\kappa_z d}. \tag{2.9}$$

Similarly, the fourth term in Eq. (2.8) can be written as

$$\mathrm{Tr}\left(\frac{k_0'}{k_0}\right)^2 \left[LR_2'L\frac{1}{D_{12}}R_1^{\dagger}\frac{1}{D_{12}^{\dagger}}\right]e^{-2\kappa_z d} = \mathrm{Tr}\left(\frac{k_0'}{k_0}\right)^2 \left[L\frac{1}{D_{21}}R_2'\frac{1}{D_{21}^{\dagger}}LR_1^{\dagger}\right]e^{-2\kappa_z d}. \tag{2.10}$$

In the following, we use the identities

$$\begin{aligned}\frac{1}{D_{12}} &= \left(\frac{k_0}{k_0'}\right)^2\left[I + R_1L\frac{1}{D_{21}}R_2'Le^{-2\kappa_z d}\right],\\ \frac{1}{D_{21}} &= \left(\frac{k_0}{k_0'}\right)^2\left[I + R_2'L\frac{1}{D_{12}}R_1Le^{-2\kappa_z d}\right].\end{aligned} \tag{2.11}$$

The second term in Eq. (2.8) can be written as

$$\begin{aligned}&\mathrm{Tr}\left(\frac{k_0'}{k_0}\right)^2 \left[LR_2'^{\dagger}L\frac{1}{D_{12}}R_1^{\dagger}\frac{1}{D_{12}^{\dagger}}\right]e^{-2\kappa_z d}\\ &= \mathrm{Tr}\left[LR_1^{\dagger}LR_2'^{\dagger}\frac{1}{D_{21}^{\dagger}} + LR_1L\frac{1}{D_{21}}R_2'LR_1^{\dagger}LR_2'^{\dagger}\frac{1}{D_{21}^{\dagger}}e^{-2\kappa_z d}\right]e^{-2\kappa_z d}.\end{aligned} \tag{2.12}$$

The first term in Eq. (2.12) can be further proceeded as

$$\mathrm{Tr}\left[LR_1^{\dagger}LR_2'^{\dagger}\frac{1}{D_{21}^{\dagger}}\right]e^{-2\kappa_z d} = \mathrm{Tr}\left[\left(\frac{k_0'}{k_0}\right)^2\frac{1}{D_{21}^{\dagger}} - I\right]. \tag{2.13}$$

The second term in Eq. (2.12) is written as

$$\mathrm{Tr}\left[LR_1L\frac{1}{D_{21}}R_2'LR_1^\dagger LR_2'^\dagger\frac{1}{D_{21}^\dagger}e^{-2\kappa_z d}\right]e^{-2\kappa_z d}$$
$$= \mathrm{Tr}\left(\frac{k_0'}{k_0}\right)\left[LR_1L\frac{1}{D_{21}}R_2'\frac{1}{D_{21}^\dagger}\right]e^{-2\kappa_z d} - \mathrm{Tr}\left[LR_1L\frac{1}{D_{21}}R_2'\right]\ e^{-2\kappa_z d}. \tag{2.14}$$

Thus, summarizing Eqs. (2.13) and (2.14), Eq. (2.12) becomes

$$\mathrm{Tr}\left[\left(\frac{k_0'}{k_0}\right)^2\frac{1}{D_{21}^\dagger}-I\right]+\mathrm{Tr}\left(\frac{k_0'}{k_0}\right)\left[LR_1L\frac{1}{D_{21}}R_2'\frac{1}{D_{21}^\dagger}\right]e^{-2\kappa_z d}-\mathrm{Tr}\left[LR_1L\frac{1}{D_{21}}R_2'\right]\ e^{-2\kappa_z d}$$
$$=\mathrm{Tr}\left[\left(\frac{k_0'}{k_0}\right)^2\frac{1}{D_{21}^\dagger}-I\right]+\mathrm{Tr}\left(\frac{k_0'}{k_0}\right)\left[LR_1L\frac{1}{D_{21}}R_2'\frac{1}{D_{21}^\dagger}\right]e^{-2\kappa_z d}$$
$$-\mathrm{Tr}\left[\left(\frac{k_0'}{k_0}\right)^2\frac{1}{D_{21}}-I\right]\ e^{-2\kappa_z d}. \tag{2.15}$$

The third term in Eq. (2.8) is

$$\mathrm{Tr}\left(\frac{k_0'}{k_0}\right)^2\left[LR_2'L\frac{1}{D_{12}}R_1\frac{1}{D_{12}^\dagger}\right]e^{-2\kappa_z d}$$
$$=\mathrm{Tr}\left[\left(\frac{k_0{}'}{k_0}\right)^2\frac{1}{D_{21}}-I\right]+\mathrm{Tr}\left(\frac{k_0'}{k_0}\right)^2\left[\frac{1}{D_{21}}R_2'^\dagger\frac{1}{D_{21}^\dagger}LR_1^\dagger L\right]e^{-2\kappa_z d}$$
$$-\mathrm{Tr}\left[\left(\frac{k_0'}{k_0}\right)^2\frac{1}{D_{21}^\dagger}-I\right], \tag{2.16}$$

Combining Eqs. (2.9), (2.10), (2.15), and (2.16), we finally arrive at

$$\tau_{1\to 2}^{evan} = \mathrm{Tr}\begin{bmatrix} LR_1L\frac{1}{D_{21}}R_2'^{\dagger}\frac{1}{D_{21}^{\dagger}} + L\frac{1}{D_{21}}R_2'\frac{1}{D_{21}^{\dagger}}LR_1^{\dagger} \\ -LR_1L\frac{1}{D_{21}}R_2'\frac{1}{D_{21}^{\dagger}} - \frac{1}{D_{21}}R_2'^{\dagger}\frac{1}{D_{21}^{\dagger}}LR_1^{\dagger}L \end{bmatrix} e^{-2\kappa_z d}$$

$$= \mathrm{Tr}\left(\frac{k_0'}{k_0}\right)^2 \left[L\left(R_1^{\dagger} - R_1\right)L\frac{1}{D_{21}}\left(R_2' - R'^{\dagger}_2\right)\frac{1}{D_{21}^{\dagger}}\right] e^{-2\kappa_z d}$$

$$= \tau_{2\to 1}^{evan}. \quad (2.17)$$

Therefore, we proved the equality for both propagative and evanescent waves. Note that we did not assume any constraint in the derivation and thus the result is valid in general in the fluctuational electrodynamics framework.

## 3. Reciprocity condition

When slab 1 satisfies the Lorentz reciprocity, the reflection matrix of the material satisfies [1]

$$R_1(\omega, -\boldsymbol{q}) = \sigma_z R_1^T(\omega, \boldsymbol{q})\sigma_z, \quad (3.1)$$

where $\sigma_z = \begin{bmatrix} 1 & 0 \\ 0 & -1 \end{bmatrix}$ is the Pauli matrix. For slab 2, the reflection matrix in the co-moving frame satisfies

$$R_2(\omega', -\boldsymbol{q}'; +V) = \sigma_z R_2^T(\omega', \boldsymbol{q}'; +V)\sigma_z. \quad (3.2)$$

For the reflection matrix of slab 2, we added an argument $V$ to indicate the direction of the Lorentz boost in the *x*-direction. The reflection matrix is written more specifically as

$$R_2(\omega', \boldsymbol{q}'; +V) = R_2\big(\gamma(\omega - q_x V), \gamma(q_x - \beta k_0), q_y, k_z; +V\big). \quad (3.3)$$

Then, flipping the sign of $\boldsymbol{q}$ results in

$$
\begin{aligned}
&R_2'\big(\gamma(\omega+q_xV),\gamma(-q_x-\beta k_0),-q_y,k_z;+V\big)\\
&=R_2'\big(\gamma\big(\omega-q_x(-V)\big),-\gamma(q_x+(-\beta)k_0),-q_y,k_z;+V\big)\\
&=R_2'(\omega',-\boldsymbol{q}';-V)=\sigma_z R'^{T}_{2}(\omega',\boldsymbol{q}';-V)\sigma_z. \qquad (3.4)
\end{aligned}
$$

The matrix of the Lorentz transformation Eq. (1.1.27) transforms under the flipping the sign of $\boldsymbol{q}$ as

$$
L(\omega,-\boldsymbol{q};+V)=L(\omega,\boldsymbol{q};-V), \qquad (3.5)
$$

$$
\sigma_z L(\omega,\boldsymbol{q};-V)\sigma_z=L^T(\omega,\boldsymbol{q};-V). \qquad (3.6)
$$

From here, we omit the frequency argument. Then, we can find the transformation of $\tilde{R}_2(\omega,\boldsymbol{q},V)$ under $\boldsymbol{q}\to-\boldsymbol{q}$ as

$$
\begin{aligned}
\tilde{R}_2(\omega,-\boldsymbol{q};+V)&=L(\omega,-\boldsymbol{q};+V)\sigma_z R'^{T}_{2}(\omega',\boldsymbol{q}';-V)\sigma_z\big(L^T(\omega,-\boldsymbol{q};+V)\big)^{-1}\\
&=\sigma_z L^T(\omega,\boldsymbol{q};-V)R'^{T}_{2}(\omega',\boldsymbol{q}';-V)L^{-1}(\omega,\boldsymbol{q};-V)\sigma_z\\
&=\sigma_z\left[\big(L^T(\omega,\boldsymbol{q};-V)\big)^{-1}R_2'(\omega',\boldsymbol{q}';-V)L(\omega,\boldsymbol{q};-V)\right]^T\sigma_z\\
&=\sigma_z\tilde{R}^T{}_2(\omega,\boldsymbol{q};-V)\sigma_z, \qquad (3.7)
\end{aligned}
$$

where in the second line, we used Eqs. (3.5), (3.6), (1.1.29) as well as $\sigma_z\sigma_z=I$.

We first consider the propagative waves. The matrices inside the trace transform as follows.

$$
\begin{aligned}
&I-\tilde{R}_2^\dagger(\omega,-\boldsymbol{q};V)\tilde{R}_2(\omega,-\boldsymbol{q};V)\\
&=I-\sigma_z\tilde{R}^*{}_2(\omega,\boldsymbol{q};-V)\tilde{R}^T{}_2(\omega,\boldsymbol{q};-V)\sigma_z\\
&=\sigma_z\big[I-\tilde{R}_2(\omega,\boldsymbol{q};-V)\tilde{R}_2^\dagger(\omega,\boldsymbol{q};-V)\big]^T\sigma_z. \qquad (3.8)
\end{aligned}
$$

$$
\begin{aligned}
\frac{1}{D_{12}(\omega,-\boldsymbol{q};V)}&=\frac{1}{I-R_1(\omega,-\boldsymbol{q})\tilde{R}_2(\omega,-\boldsymbol{q};V)e^{2ik_zd}}\\
&=\frac{1}{I-\sigma_z R_1^T(\omega,\boldsymbol{q})\tilde{R}^T{}_2(\omega,\boldsymbol{q};-V)\sigma_z e^{2ik_zd}}\\
&=\sigma_z\frac{1}{D_{21}^T(\omega,\boldsymbol{q};-V)}\sigma_z. \qquad (3.9)
\end{aligned}
$$

Using Eqs. (3.8) and (3.9), it is straightforward to show that

$$\tau_{1\to 2}^{prop}(\omega, -\boldsymbol{q}; V) = \tau_{2\to 1}^{prop}(\omega, \boldsymbol{q}; -V). \tag{3.10}$$

It is also straightforward to show that the above identity holds for the evanescent waves. Thus, we obtain the constraint on the transmission coefficient when the slabs satisfy the Lorentz reciprocity

$$\tau_{1\to 2}(\omega, -\boldsymbol{q}; V) = \tau_{2\to 1}(\omega, \boldsymbol{q}; -V). \tag{3.11}$$

The result shows that the transmission coefficient can exhibit non-reciprocity as far as $V \neq 0$ even when the materials are composed of reciprocal materials.

## 4. Proof of Eq. (24)

We provide the proof of the identity Eq. (24)

$$T_1 f_{x,2} V \leq \left| T_1 - \frac{T_2'}{\gamma} \right| \operatorname{sgn}\left( T_1 - \frac{T_2'}{\gamma} \right) \varphi_{1\to 2}, \tag{4.1}$$

in the framework of fluctuational electrodynamics. We introduce a shorthand notation $x = (\omega, \boldsymbol{q})$. For an arbitrary real scalar function $g(x)$ and complex vector $\boldsymbol{P}(x)$, we consider the following integral

$$\int dx\, g^2(x) \boldsymbol{P}^\dagger(x) \omega \big( I - R_1(x) R_1^\dagger(x) \big) \boldsymbol{P}(x), \tag{4.2}$$

where $R_1$ is the reflection matrix of slab 1. This can be rewritten as:

$$\begin{aligned} &\int dx\, g^2(x) \boldsymbol{P}^\dagger(x) \omega \big( I - R_1(x) R_1^\dagger(x) \big) \boldsymbol{P}(x) \\ &= \int dx \int dx'\, g(x) g(x') \boldsymbol{P}^\dagger(x) \omega \left( I - R_1(x) R_1^\dagger(x') \right) \delta(x - x') \boldsymbol{P}(x'). \end{aligned} \tag{4.3}$$

For the propagative waves $q < \frac{\omega}{c}$, the fluctuation-dissipation theorem is

$$\langle \boldsymbol{E}(x)\boldsymbol{E}^{\dagger}(x')\rangle_S = \pi(2\pi)^2\sqrt{\frac{\mu_0}{\varepsilon_0}\frac{k_0}{k_z}}\Theta(\omega,T_1)\big(I - R_1R_1^{\dagger}\big)\delta(x-x'). \tag{4.4}$$

Noting that $k_z > 0$ for the propagative waves and $\Theta(\omega,T_1) > 0$ for $\omega > 0$, Eq. (4.3) becomes:

$$\begin{aligned}
&\int dx \int dx'\, g(x)g(x')\boldsymbol{P}^{\dagger}(x)\omega\left(I - R_1(x)R_1^{\dagger}(x')\right)\delta(x-x')\boldsymbol{P}(x') \\
&= \frac{ck_z}{\pi(2\pi)^2\Theta(\omega,T_1)}\sqrt{\frac{\varepsilon_0}{\mu_0}}\int dx \int dx'\, g(x)g(x')\boldsymbol{P}^{\dagger}(x)\langle \boldsymbol{E}(x)\boldsymbol{E}^{\dagger}(x')\rangle_S\boldsymbol{P}(x') \\
&= \frac{ck_z}{\pi(2\pi)^2\Theta(\omega,T_1)}\sqrt{\frac{\varepsilon_0}{\mu_0}}\langle\left(\int dx\, g(x)\boldsymbol{E}^{\dagger}(x)\boldsymbol{P}(x)\right)^{\dagger}\left(\int dx'\, g(x')\boldsymbol{E}^{\dagger}(x')\boldsymbol{P}(x')\right)\rangle_S \geq 0.
\end{aligned} \tag{4.5}$$

Thus, Eq. (4.2) is positive

$$\int dx\, g^2(x)\boldsymbol{P}^{\dagger}(x)\omega\left(I - R_1(x)R_1^{\dagger}(x)\right)\boldsymbol{P}(x) \geq 0. \tag{4.6}$$

If this is positive for arbitrary real scalar function $g(x)$, the integrand is positive for each $x$. Thus,

$$\boldsymbol{P}^{\dagger}(\omega,\boldsymbol{q})\omega\left(I - R_1(\omega,\boldsymbol{q})R_1^{\dagger}(\omega,\boldsymbol{q})\right)\boldsymbol{P}(\omega,\boldsymbol{q}) \geq 0 \quad \forall \boldsymbol{P}. \tag{4.7}$$

Thus, $\omega\left(I - R_1(\omega,\boldsymbol{q})R_1^{\dagger}(\omega,\boldsymbol{q})\right)$ is positive semidefinite. Following the same argument, for the evanescent waves, we have

$$\boldsymbol{P}^{\dagger}(\omega,\boldsymbol{q})\omega\frac{\left(R_1(\omega,\boldsymbol{q}) - R_1^{\dagger}(\omega,\boldsymbol{q})\right)}{2i}\boldsymbol{P}(\omega,\boldsymbol{q}) \geq 0 \quad \forall \boldsymbol{P}. \tag{4.8}$$

Similarly, the same argument for the reflection matrices about slab 2 gives

$$\boldsymbol{P}^\dagger(\omega',\boldsymbol{q}')\omega'\left(I - R'_2(\omega',\boldsymbol{q}')R'^\dagger_2(\omega',\boldsymbol{q})\right)\boldsymbol{P}(\omega',\boldsymbol{q}) \geq 0 \quad \forall \boldsymbol{P}, \tag{4.9}$$

$$\boldsymbol{P}^\dagger(\omega',\boldsymbol{q}')\omega'\frac{\left(R'_2(\omega',\boldsymbol{q}') - R'^\dagger_2(\omega',\boldsymbol{q}')\right)}{2i}\boldsymbol{P}(\omega',\boldsymbol{q}') \geq 0 \quad \forall \boldsymbol{P}. \tag{4.10}$$

The propagative waves in the rest frame are also the propagative waves in the co-moving frame, i.e., if $q < \frac{\omega}{c}$ then $q' < \frac{\omega'}{c}$. Then by using the properties of the $L$ matrix for the propagative waves, we have

$$I - \tilde{R}_2\tilde{R}_2^\dagger = \left(\frac{k_0}{k'_0}\right)^2 L\left(I - R'_2R'^\dagger_2\right)L^\dagger. \tag{4.11}$$

Thus, since $\omega'\left(I - R'_2R'^\dagger_2\right)$ is positive semidefinite, $\omega'\left(I - \tilde{R}_2\tilde{R}_2^\dagger\right)$ is also positive semidefinite because for arbitrary vector $\boldsymbol{v} \in \mathbb{C}$, we have $\boldsymbol{v}^\dagger BAB^\dagger\boldsymbol{v} \geq 0$ if $A$ is positive semidefinite. Let us write $A = I - \tilde{R}_2\tilde{R}_2^\dagger$ and $B = I - \tilde{R}_2^\dagger\tilde{R}_2$. $\omega' A$ is positive semidefinite as shown above. We have

$$\omega'\tilde{R}_2B\tilde{R}_2^\dagger = \omega'\tilde{R}_2\tilde{R}_2^\dagger - \omega'\tilde{R}_2\tilde{R}_2^\dagger\tilde{R}_2\tilde{R}_2^\dagger = \omega'\tilde{R}_2\tilde{R}_2^\dagger A = \omega' A\tilde{R}_2\tilde{R}_2^\dagger. \tag{4.12}$$

Since $\omega' A$ and $\tilde{R}_2\tilde{R}_2^\dagger$ are respectively positive semidefinite and they commute, the product of these is also positive semidefinite. Then $\omega'\tilde{R}_2B\tilde{R}_2^\dagger$ is positive semidefinite. From the same argument above, this means $\omega' B = \omega'\left(I - \tilde{R}_2^\dagger\tilde{R}_2\right)$ is positive semidefinite.

The transmission coefficient of the propagative waves is

$$\omega\omega'\tau_{1\to2}^{prop}(\omega,\boldsymbol{q};V) = \omega\omega'\mathrm{Tr}\left[\left(I - \tilde{R}_2^\dagger\tilde{R}_2\right)\frac{1}{D_{12}}\left(I - R_1R_1^\dagger\right)\frac{1}{D_{12}^\dagger}\right]. \tag{4.13}$$

Because $\omega'\left(I - \tilde{R}_2^\dagger\tilde{R}_2\right)$ and $\frac{1}{D_{12}}\omega\left(I - R_1R_1^\dagger\right)\frac{1}{D_{12}^\dagger}$ are positive semidefinite and the trace of the product of two positive semidefinite matrices is non-negative, we can show that

$$\omega\omega'\tau_{1\to2}^{prop}(\omega,\boldsymbol{q};V)\geq 0. \tag{4.14}$$

We can follow the same argument for the evanescent waves. Combining the propagative and evanescent waves, we prove that

$$\omega\omega'\tau_{1\to2}(\omega,\boldsymbol{q};V)\geq 0. \tag{4.15}$$

With Eq. (4.15), we are ready to prove Eq. (4.1). From the fluctuational electrodynamics formalism, we have

$$\left|T_1-\frac{T_2'}{\gamma}\right|\operatorname{sgn}\left(T_1-\frac{T_2'}{\gamma}\right)\varphi_{1\to2}-T_1f_{x,2}V$$
$$=\int_0^\infty\frac{d\omega}{2\pi}\int\frac{d\boldsymbol{q}}{2\pi}\hbar\omega\omega'\left[T_1(\omega-q_xV)-\frac{T_2'}{\gamma}\omega\right]\left[n_B(\omega,T_1)-n_B\left(\omega-q_xV,\frac{T_2'}{\gamma}\right)\right]\times$$
$$\frac{\tau_{1\to2}(\omega,\boldsymbol{q};V)}{\omega\omega'}. \tag{4.16}$$

From Eq. (4.15), $\frac{\tau_{1\to2}(\omega,\boldsymbol{q};V)}{\omega\omega'}\geq 0$. Note that only the sign of $\omega\omega'$ matters. Also, we can show that

$$\omega\omega'\left[T_1(\omega-q_xV)-\frac{T_2'}{\gamma}\omega\right]\left[n_B(\omega,T_1)-n_B\left(\omega-q_xV,\frac{T_2'}{\gamma}\right)\right]\geq 0, \tag{4.17}$$

for any frequency $\omega$ and $\omega'$. Combining these two, we prove Eq. (4.1).

## 5. Scaling behavior of radiative heat flux with respect to *V*

The radiative heat flux and the shear stress for non-reciprocal materials have non-zero linear scaling with respect to $V$. However, those for reciprocal materials scale quadratically. For small non-relativistic velocity, i.e., $\gamma\approx 1$, the expansion of the number distribution $n_B(\omega',T_2')$ with respect to the velocity is

$$\varphi_{1\to 2} \approx \int_0^\infty \frac{d\omega}{2\pi} \int_{-\infty}^\infty \frac{dq_y}{2\pi} \int_0^\infty \frac{dq_x}{2\pi} \hbar\omega \times \begin{bmatrix} -\frac{dn_B(\omega, T_2')}{dV}\Big|_{V=0} q_x V\{\tau_{1\to 2}(q_x, q_y; V, B) - \tau_{1\to 2}(-q_x, q_y; V, B)\} \\ +\Delta n_B^0\{\tau_{1\to 2}(q_x, q_y; V, B) + \tau_{1\to 2}(-q_x, q_y; V, B)\} \end{bmatrix} + \mathcal{O}(V^2), \quad (5.1)$$

where $\Delta n_B^0 = n_B(\omega, T_1) - n_B(\omega, T_2')$ and we add $B$ as an argument to induce non-reciprocity in the system other than the Doppler shift by $V$. For reciprocal systems, i.e., $B = 0$, the lowest order of $\tau_{1\to 2}(q_x, q_y; V) \pm \tau_{1\to 2}(-q_x, q_y; V)$ with respect to $V$ is $\mathcal{O}(V)$ for $-$ and $\mathcal{O}(V^2)$ for $+$. Thus, the radiative heat flux scales with respect to the velocity as $\varphi_{1\to 2} \propto V^2$. However, for non-reciprocal systems, the first order in $\tau_{1\to 2}(q_x, q_y; V) - \tau_{1\to 2}(-q_x, q_y; V)$ is non-zero due to the asymmetry of the transmission coefficient even at $V = 0$, and the linear term with respect to $V$ is nonzero and the radiative heat flux scales as $\varphi_{1\to 2} \propto V$. This different scaling is shown in Fig. S2 where the normalized radiative heat flux as a function of velocity is plotted for the reciprocal ($B = 0$) and non-reciprocal ($B \neq 0$) material systems. Therefore, the linear scaling of the radiative heat flux with respect to $V$ in the non-reciprocal system can be considered a signature of non-reciprocal radiative heat transfer and may be feasible for experimental observations.

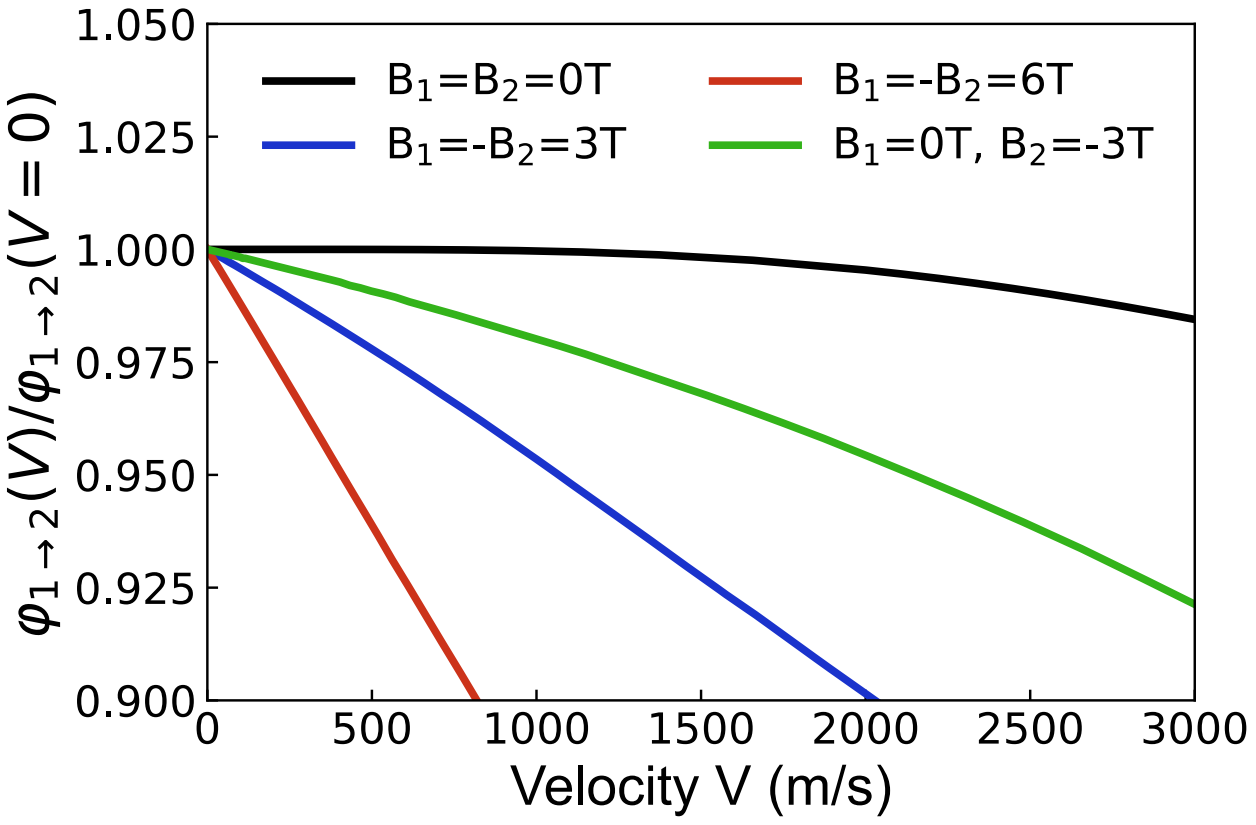

Fig. S2. Radiative heat flux at the relative velocity $V$ normalized by that at rest. Linear scaling of $\varphi_{1\to 2}$ with respect to $V$ is non-zero for non-reciprocal materials. The scaling is quadratic for reciprocal materials.

## 6. Lorentz transformation of radiative heat flux and shear stress

In this section, we derive the radiative heat flux in the co-moving frame and show that the relation between the radiative heat flux and the shear stress in the rest and co-moving frames are related by the Lorentz boost in the $x$-direction if we define the four-vector as $f^{\mu} = (-\frac{\varphi_{1\to 2}}{c}, f_{x,1}, f_{y,1}, f_{z,1})$.

First, we give an argument as to why the four-vector in our system is given as above. First we consider the local conservation law of energy given as [2]

$$\frac{\partial u}{\partial t} + \nabla \cdot \boldsymbol{S} = -\boldsymbol{J} \cdot \boldsymbol{E}, \tag{6.1}$$

where $u$ is the electromagnetic energy density, $\boldsymbol{S}$ is the Poynting vector and $\boldsymbol{J}$ is the current density. In our system, we take the ensemble average of Eq. (6.1) since the electromagnetic fields and charges are fluctuating, and the time variation of the electromagnetic energy is zero. Integrating Eq. (6.1) over the volume of slab 1, we have

$$-\int dAS_z = \int \boldsymbol{J} \cdot \boldsymbol{E}\, dV\,. \tag{6.2}$$

The left-hand side of Eq. (6.2) is the energy coming into slab 1 through the surface $d\boldsymbol{A} = dAe_z$ and is equal to the right-hand side which is the work done on the charged particles constituting slab 1 by the electromagnetic fields. Thus, divided by the surface area $A$, we have

$$-\varphi_{1\to 2} = \frac{1}{A}\int \boldsymbol{J} \cdot \boldsymbol{E}\, dV. \tag{6.3}$$

Similarly, we consider the local conservation law of linear momentum

$$\boldsymbol{f}_{mech} + \varepsilon_0 \frac{\partial}{\partial t}(\boldsymbol{E} \times \boldsymbol{H}) = \nabla \cdot \tau, \tag{6.4}$$

where $\boldsymbol{f}_{mech} = \rho\boldsymbol{E} + \boldsymbol{J} \times \boldsymbol{E}$ is the force density on charges due to the electromagnetic fields, $\varepsilon_0 \frac{\partial}{\partial t}(\boldsymbol{E} \times \boldsymbol{H})$ is the electromagnetic field momentum, and $\tau$ is the Maxwell stress tensor. Since the electromagnetic fields and charges are fluctuating, we take the ensemble average. The term of the time-derivative of the electromagnetic momentum is zero. Integrating over the volume of slab 1 and dividing by the area, we have

$$\boldsymbol{f} = \frac{1}{A}\int dA\boldsymbol{n}_{\boldsymbol{z}} \cdot \tau = \frac{1}{A}\int \boldsymbol{f}_{mech}\, dV, \tag{6.5}$$

where $\boldsymbol{f}$ is the shear stress and radiation pressure defined in Eq. (1.3.2).

The field strength tensor is

$$F^{\mu\nu} = \begin{bmatrix} 0 & -\frac{E_x}{c} & -\frac{E_y}{c} & -\frac{E_z}{c} \\ \frac{E_x}{c} & 0 & -B_z & B_y \\ \frac{E_y}{c} & B_z & 0 & -B_x \\ \frac{E_z}{c} & -B_y & B_x & 0 \end{bmatrix}, \tag{6.6}$$

and the four-current is $j^{\mu} = (c\rho, j_x, j_y, j_z)$. From $F^{\mu\nu}$ and $j^{\mu}$, we can show that

$$f^{\mu} = F^{\mu\nu} j_{v}, \tag{6.7}$$

where $f^{\mu} = (\frac{\boldsymbol{j}\cdot\boldsymbol{E}}{c}, f_{mech,x}, f_{mech,y} f_{mech,z})$. Thus, $f^{\mu}$ is a four-vector.

We restrict ourselves to the Lorentz boosts only in $x$ or $y$ direction. In this case, $\frac{1}{A}\int dV$ is Lorentz invariant. Then, by integrating $f^{\mu}$ over the volume of slab 1 and dividing it by the surface area and using Eqs. (6.3) and (6.5), $(-\frac{\varphi_{1\to 2}}{c}, f_{x,1}, f_{y,1}, f_{z,1})$ is also a four-vector. Then for the Lorentz boost in the $x$ direction gives

$$\begin{bmatrix} -\frac{\varphi'_{1\to 2}}{c} \\ f'_{x,1} \\ f'_{y,1} \\ f'_{z,1} \end{bmatrix} = \begin{bmatrix} \gamma & -\gamma\beta & 0 & 0 \\ \gamma\beta & \gamma & 0 & 0 \\ 0 & 0 & 1 & 0 \\ 0 & 0 & 0 & 1 \end{bmatrix} \begin{bmatrix} -\frac{\varphi_{1\to 2}}{c} \\ f_{x,1} \\ f_{y,1} \\ f_{z,1} \end{bmatrix}. \tag{6.8}$$

The first row gives Eq. (8) in the main text: $\varphi'_{1\to 2} = \gamma\varphi_{1\to 2} + \gamma f_{x,1}V$. Note that if Lorentz boost is in the *z* direction, then $(-\frac{\varphi_{1\to 2}}{c}, f_{x,1}, f_{y,1}, f_{z,1})$ is not four-vector.

Next, we derive Eq. (8) in the main text directly. Similar to Eq. (1.2.5), the *z*-component of the Poynting flux vector in the co-moving frame is

$$\varphi'_z = \int_0^\infty \frac{d\omega'}{2\pi} \int \frac{d\boldsymbol{q}'}{(2\pi)^4} \frac{1}{\pi} \langle \boldsymbol{E}'(\boldsymbol{q}', z', \omega') \times \boldsymbol{H}'^*(\boldsymbol{q}', z', \omega') \rangle \cdot \boldsymbol{e}_{z'}|_{z'=0}, \tag{6.9}$$

where all the quantities are defined in the co-moving frame. Using Eqs. (1.1.17) and (1.1.19), we obtain

$$\begin{aligned} &\langle \boldsymbol{E}'(\boldsymbol{q}', z', \omega') \times \boldsymbol{H}'^*(\boldsymbol{q}', z', \omega') + \boldsymbol{E}'^*(\boldsymbol{q}', z', \omega') \times \boldsymbol{H}'(\boldsymbol{q}', z', \omega') \rangle_E \cdot \boldsymbol{e}_{z'}|_{z'=0} \\ &= \frac{2}{c\mu_0} \mathrm{Tr} \begin{bmatrix} \frac{\mathrm{Re}[k_z]}{k'_0} \left\{ \langle \begin{bmatrix} v'_s \\ v'_p \end{bmatrix} \begin{bmatrix} v'^*_s & v'^*_p \end{bmatrix} \rangle_S - \langle \begin{bmatrix} w'_s \\ w'_p \end{bmatrix} \begin{bmatrix} w'^*_s & w'^*_p \end{bmatrix} \rangle_S \right\} \\ +i \frac{\mathrm{Im}[k_z]}{k'_0} \left\{ \langle \begin{bmatrix} v'_s \\ v'_p \end{bmatrix} \begin{bmatrix} w'^*_s & w'^*_p \end{bmatrix} \rangle_S - \langle \begin{bmatrix} w'_s \\ w'_p \end{bmatrix} \begin{bmatrix} v'^*_s & v'^*_p \end{bmatrix} \rangle_S \right\} \end{bmatrix}. \end{aligned} \tag{6.10}$$

Note that the symmetrized correlation function is with respect to the material Hamiltonian in the co-moving frame. Consider the propagative waves first, i.e., $\mathrm{Im}[k_z] = 0$. By using Eq. (1.1.26) and the cyclic property of the trace, the first term in Eq. (6.10) becomes

$$\frac{2}{c\mu_0} \mathrm{Tr} \frac{\mathrm{Re}[k_z]}{k'_0} \left\{ \langle L^\dagger L \begin{bmatrix} v_s \\ v_p \end{bmatrix} \begin{bmatrix} v_s^* & v_p^* \end{bmatrix} \rangle_S - (L^T)^\dagger L^T \langle \begin{bmatrix} w_s \\ w_p \end{bmatrix} \begin{bmatrix} w_s^* & w_p^* \end{bmatrix} \rangle_S \right\}. \tag{6.11}$$

For the propagative waves, $L^{\dagger}L = L^{T}L = \left(\frac{k_0'}{k_0}\right)^2$ and $(L^T)^{\dagger}L^T = LL^T = \left(\frac{k_0'}{k_0}\right)^2$, and Eq. (6.11) becomes

$$\frac{k_0'}{k_0}\frac{2}{c\mu_0}\mathrm{Tr}\frac{\mathrm{Re}[k_z]}{k_0}\left\{\langle\begin{bmatrix} v_s \\ v_p \end{bmatrix}\begin{bmatrix} v_s^* & v_p^* \end{bmatrix}\rangle_S - \langle\begin{bmatrix} w_s \\ w_p \end{bmatrix}\begin{bmatrix} w_s^* & w_p^* \end{bmatrix}\rangle_S\right\}. \tag{6.12}$$

This is the same result as Eq. (1.2.6) multiplied by the factor $\frac{k_0'}{k_0}$. We can similarly show the same result for the evanescent components. Therefore, the radiative heat flux from slabs 1 to 2 measured in the co-moving frame is expressed as

$$\varphi_{1\to2}' = \int_0^{\infty}\frac{d\omega'}{2\pi}\int\frac{d\boldsymbol{q}'}{(2\pi)^2}\frac{k_0'}{k_0}\left[\frac{\hbar\omega}{e^{\frac{\hbar\omega}{k_B T_1}}-1} - \frac{\hbar\omega}{e^{\frac{\hbar\gamma(\omega-q_xV)}{k_B T_2'}}-1}\right]\tau_{1\to2}(\omega,\boldsymbol{q};V). \tag{6.13}$$

Since $k_z = k_z'$ and $d\omega dq_x dq_y dk_z$ is Lorentz invariant, we have $d\omega dq_x dq_y = d\omega' dq_x' dq_y'$. By using the Lorentz transformation of the angular frequency Eq. (1.1.8), we have

$$\begin{aligned}\varphi_{1\to2}' &= \int_0^{\infty}\frac{d\omega}{2\pi}\int\frac{d\boldsymbol{q}}{(2\pi)^2}\gamma(1-\frac{\beta}{k_0}q_x)\left[\frac{\hbar\omega}{e^{\frac{\hbar\omega}{k_B T_1}}-1} - \frac{\hbar\omega}{e^{\frac{\hbar\gamma(\omega-q_xV)}{k_B T_2'}}-1}\right]\tau_{1\to2}(\omega,\boldsymbol{q};V) \\ &= \gamma\varphi_{1\to2} + \gamma f_{x,1}V.\end{aligned} \tag{6.14}$$

## 7. Plot of transmission coefficient and radiative heat flux for other $q$ directions

Figure S3 shows the plot of the transmission coefficient $\tau$ as a function of the frequency $\omega$ and $x$-component of the wavevector $q_x$ at $V = 1.4 \times 10^3$ m/s for the three wave propagation directions: $(q_x, q_y) = (q_x, 0)$, $(q_x, q_y) = (q_x, q_x)$, i.e., $q = \sqrt{q_x^2 + q_y^2}, \phi = \frac{\pi}{4}, \frac{3\pi}{4}$ in the polar coordinates, and $(q_x, q_y) = (0, q_y)$. From panel (a), we can see that the transmission coefficient is almost identical to that at $V = 0$ m/s, which allows us to ignore the velocity-dependence of the transmission coefficient in the heat engine regime in Fig. (2) in the main text. For the waves propagating along the $y$ direction as shown in panel (c), the transmission coefficient is reciprocal since the waves propagate along the direction of the external static magnetic fields, i.e., the Faraday configuration.

Figures S4 and S5 show the plot of the transmission coefficient $\tau$ as a function of the frequency $\omega$ and $x$-component of the wavevector $q_x$ at $V = 10^4$ m/s and $V = 1.5 \times 10^5$ m/s. The wave propagation directions are $(q_x, q_y) = (q_x, q_x)$, i.e., $q = \sqrt{q_x^2 + q_y^2}, \phi = \frac{\pi}{4}, \frac{3\pi}{4}$ in the polar coordinates and $(q_x, q_y) = (0, q_y)$ in Figs. S4 and S5, respectively. For the waves propagating along the $y$ direction as shown in Fig. S5, the transmission coefficient and the radiative heat flux is reciprocal since the waves propagate along the direction of the external static magnetic fields, i.e., the Faraday configuration.

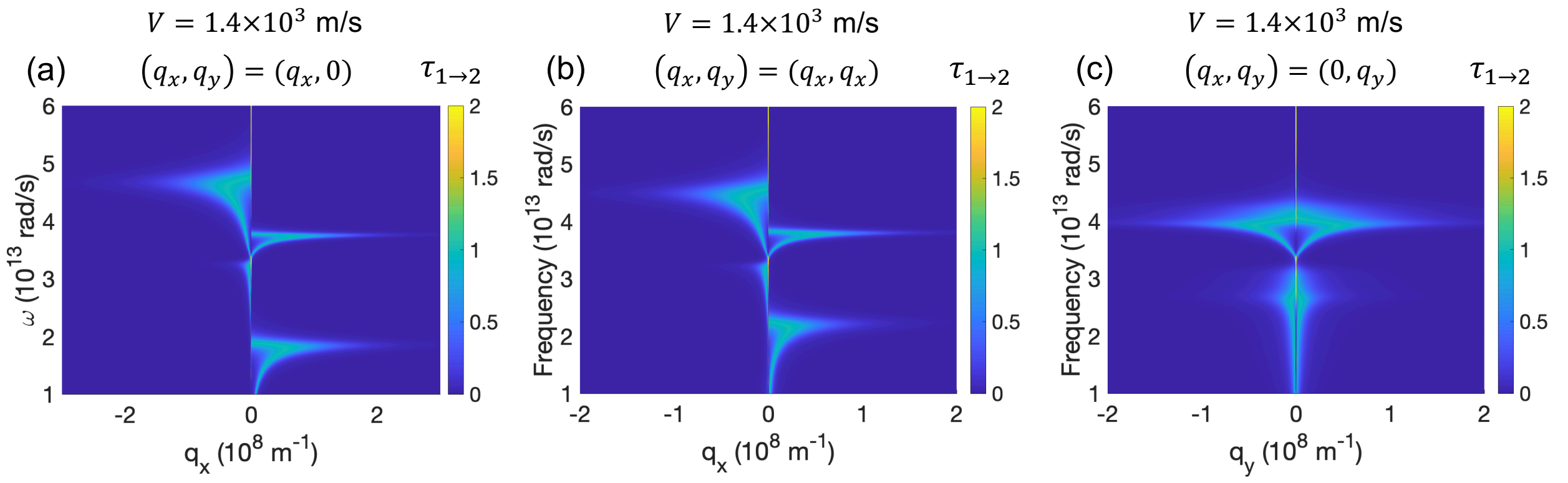

Fig. S3. Plot of the transmission coefficient $\tau$ as a function of the frequency $\omega$ at $V = 1.4 \times 10^3$ m/s for the in-plane wavevector components of (a) $\boldsymbol{q} = (q_x, 0)$, (b) $\boldsymbol{q} = (q_x, q_x)$, and (c) $\boldsymbol{q} = (0, q_y)$.

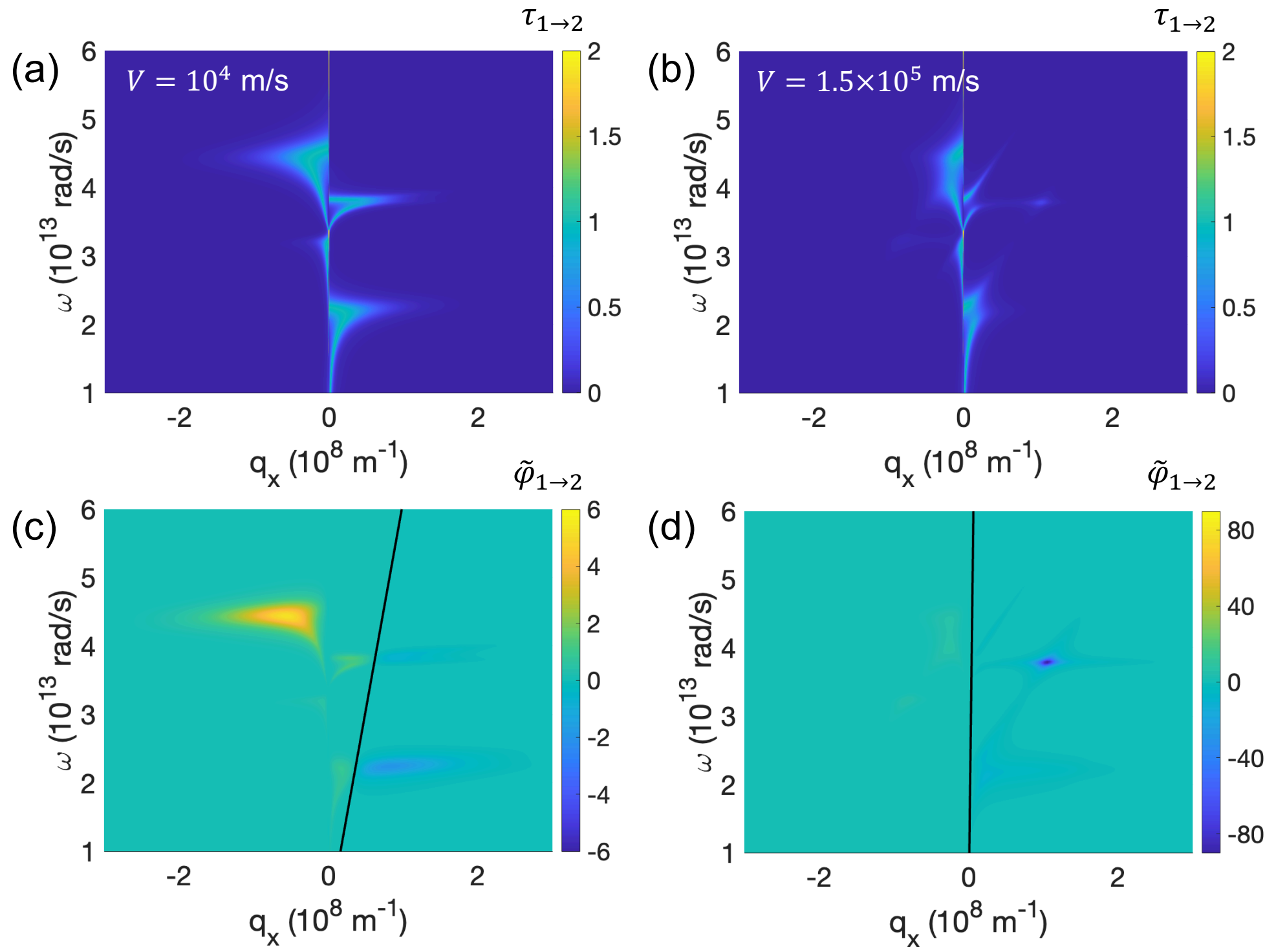

Fig. S4. (a-c) Plot of the transmission coefficient $\tau$ as a function of the frequency $\omega$ and $x$-component of the wavevector $q_x$ at $V = 10^4$ m/s and $V = 1.5 \times 10^5$ m/s. (d-f): Plot of mode-resolved radiative heat flux $\tilde{\varphi}_{1\to2}(\omega, q_x, q_y)$ defined as $\varphi_{1\to2} = \int_0^\infty d\omega \int_{-\infty}^{\infty} d\boldsymbol{q}\, \tilde{\varphi}_{1\to2}(\omega, q_x, q_y)$. The unit is $[10^{-15}\,\frac{\mathrm{W\cdot s}}{\mathrm{rad}}]$. The physical parameters are the same as those in Fig. 2 in the main text. The black lines panels (e) and (f) are the cooling condition in Eq. (22) in the main text. For all the panels, $(q_x, q_y) = (q\cos\phi, q\sin\phi)$ where $q = \sqrt{q_x^2 + q_y^2}$, and $\phi = \frac{\pi}{4}, \frac{3\pi}{4}$.

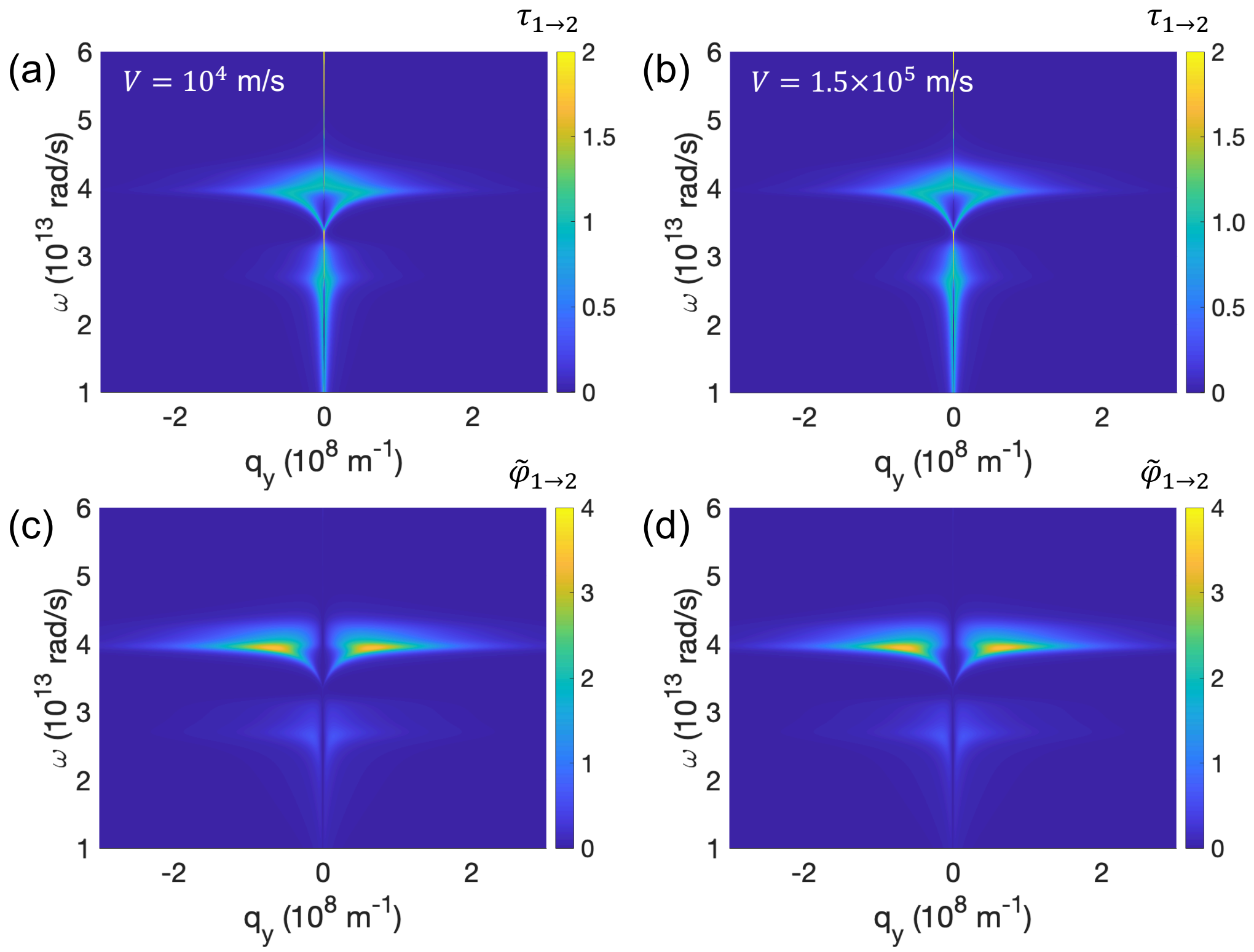

Fig. S5. (a-b) Plot of the transmission coefficient $\tau$ as a function of the frequency $\omega$ and $x$-component of the wavevector $q_y$ at $V = 10^4$ m/s and $V = 1.5 \times 10^5$ m/s. (c-d): Plot of the mode-resolved radiative heat flux $\tilde{\varphi}_{1\to2}(\omega, q_x, q_y)$ defined in the caption of Fig. S4. The physical parameters are the same as those in Fig. 2 in the main text. For all the panels, $(q_x, q_y) = (0, q_y)$.

## 8. Heat pump by magnetized plasma

Figure S6 (a) shows the radiative heat flux between two magnetized plasma as a function of the relative velocity. The dielectric function of slabs 1 and 2 are obtained by turning off the phonon contribution in the dielectric function of InSb, i.e., $\varepsilon_{ph} = 0$ in Eq. (9) in the main text. As shown in Fig. S6 (a), the system made of magnetized plasma can operate as a heat pump. Figure S6 (b) and (c) show the transmission coefficient and the mode-resolved radiative heat flux from slabs 1 to 2 for the direction $(q_x, q_y) = (q_x, 0)$ at $V = 10^5$ m/s. The magnetic fields on the slabs are $B_1 = -B_2 = 3$T. At this condition, the system operates as a heat engine since $\varphi_{1\to2} < 0$. The contour plot shows that the coincidental resonance between the surface plasmon and phonon polaritons modes that we observe in Fig. 6 (b) and (d) in the main text is absent. Thus, the operation as a heat pump does not require such coincidental resonance. Furthermore, the negative frequency modes with $\omega' < 0$ are not necessary to achieve the heat pump.

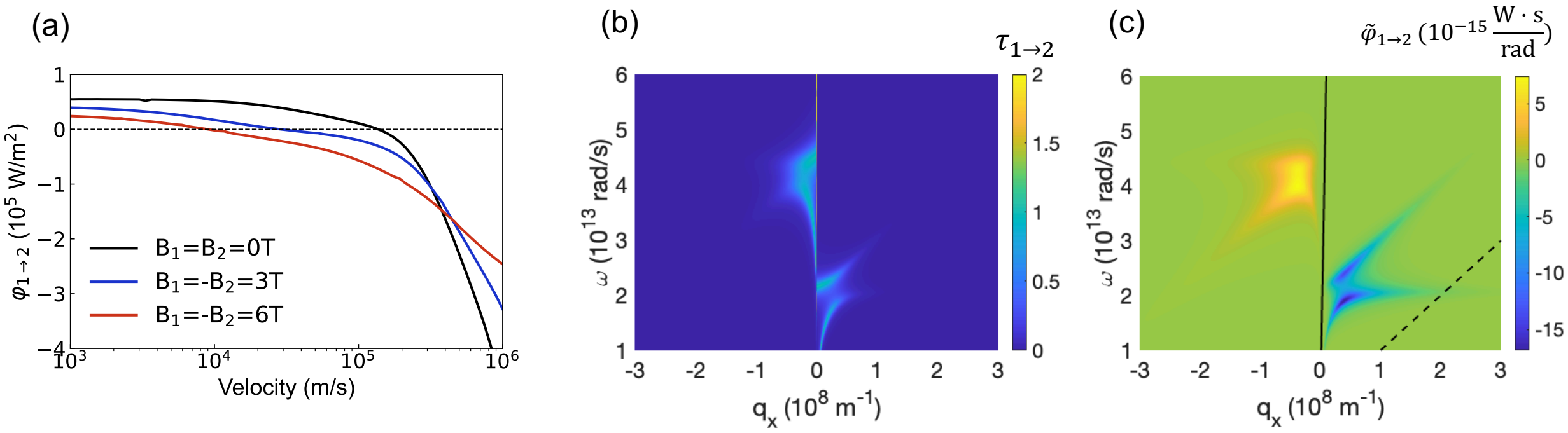

Fig. S6. (a) Radiative heat flux from slabs 1 to 2 for different magnitudes of externally applied static magnetic fields. For all panels, slabs 1 and 2 are the magnetized plasma in which the phonon contribution in the dielectric function of InSb is turned off. (b) plot of the transmission coefficient $\tau$ as a function of the frequency $\omega$ and $x$-component of the wavevector $q_x$ and (c) plot of mode-resolved radiative heat flux. For both panels (b) and (c), the modes propagateing along the $x$-direction, i.e., $\boldsymbol{q} = (q_x, 0)$ are shown and the magnetic fields on the slabs are $B_1 = -B_2 = 3$ T. All the other physical quantities are the same as those in Fig.2 in the main text. The solid and dashed lines in panel (c) are the cooling condition Eq. (22) in the main text, and the line below which the angular frequency in the co-moving frame is negative, i.e., $\omega' < 0$, respectively.